\documentclass[lettersize,journal,twocolumn]{IEEEtran}
\usepackage{amsmath,amsfonts}
\usepackage{algorithmic}
\usepackage{algorithm}
\usepackage{array}
\usepackage{textcomp}
\usepackage{picins}
\usepackage{color}
\usepackage{stfloats}
\usepackage{threeparttable}
\usepackage{url}
\usepackage[none]{hyphenat}
\usepackage{verbatim}
\usepackage{graphicx}
\usepackage{graphics} 
\usepackage{epstopdf}
\usepackage{float}  
\usepackage{subfigure}  
\usepackage{booktabs}
\usepackage{cite}
\usepackage{csquotes}
\UseRawInputEncoding
\begin{document}

\title{Joint Beamforming Design for Integrated Sensing and Communication Systems with Hybrid-Colluding Eavesdroppers}
\author{Meiding~Liu,~\IEEEmembership{Student Member,~IEEE},
        Zhengchun~Zhou,~\IEEEmembership{Member,~IEEE},
        $ \text{Qiao Shi}^{*} $,~\IEEEmembership{Member,~IEEE},\\
        Guyue Li,~\IEEEmembership{Member,~IEEE},
        Zilong Liu,~\IEEEmembership{Senior Member,~IEEE},
        Pingzhi Fan,~\IEEEmembership{Fellow,~IEEE},\\
        and
        Inkyu Lee,~\IEEEmembership{Fellow,~IEEE}
        \thanks{This work was supported in part by the National Natural Science Foundation of China under Grants U23A20274, 62350610267 and 62131016, in part by the Natural Science Foundation of Sichuan Province under Grants 2024NSFTD0015 and 2024NSFSC1418, in part by project of central government to guide local scientific and technological development under Grant 2024ZYD0012, in part by the Open Research Fund of Key Laboratory of Analytical Mathematics and Applications (Fujian Normal University), Ministry of Education, under Grant JAM2406, in part by the China Postdoctoral Science Foundation under Grant 2023M742901, in part by the National Research Foundation of Korea (NRF) funded by the Ministry of Science and ICT (MSIT), Korea Government under Grant 2022R1A5A1027646, and in part by the Institute of Information and Communications Technology Planning and Evaluation (IITP) Grant funded by the Korea Government (MSIT) and System Development of Upper-Mid Band Smart Repeater under Grant RS2024-00397480. The authors warmly thank Prof. Bin Dai for his interesting discussion and several exciting suggestions along this work. This paper was presented in part at the 2024 IEEE International Conference on Signal, Information and Data Processing \cite{conf}, Zhuhai, China, November 2024.}
        \thanks{Meiding Liu, Zhengchun Zhou and Qiao Shi are with the School of Information Science and Technology, Southwest Jiaotong University, Chengdu 611756, China, and also with the Key Laboratory of Analytical Mathematics and Applications (Fujian Normal University), Ministry of Education, P. R. China (e-mail:lmd@my.swjtu.edu.cn; zzc@swjtu.edu.cn; qiaoshi@swjtu.edu.cn). (\emph{Corresponding author: Qiao Shi}.)}
        \thanks{Guyue Li is with the School of Cyber Science and Engineering, Southeast University, Nanjing 210096, China, and also with the Purple Mountain Laboratories for Network and Communication Security, Nanjing 210096, China (e-mail: guyuelee@seu.edu.cn).}
        \thanks{Zilong Liu is with the School of Computer Science and Electronics Engineering, University of Essex, Colchester CO4 3SQ, U.K. (e-mail:zilong.liu@essex.ac.uk).}
        \thanks{Pingzhi Fan is with the Key Laboratory of Information Coding and Transmission, Southwest Jiaotong University, Chengdu 611756, China (e-mail:pzfan@swjtu.edu.cn).}
        \thanks{Inkyu Lee is with the School of Electrical Engineering, Korea University, Seoul 02841, South Korea (e-mail: inkyu@korea.ac.kr).}%
}



\maketitle

\begin{abstract}
In this paper, we consider the physical layer security (PLS) problem for integrated sensing and communication (ISAC) systems in the presence of hybrid-colluding eavesdroppers, where an active eavesdropper (AE) and a passive eavesdropper (PE) collude to intercept the confidential information. To ensure the accuracy of sensing while preventing the eavesdropping, a base station transmits a signal consisting of information symbols and sensing waveform, in which the sensing waveform can be also used as artificial noise to interfere with eavesdroppers. Under this setup, we propose an alternating optimization-based two stage scheme (AO-TSS) for improving the sensing and communication performance. In the first stage, based on the assumptions that the perfect channel state information (CSI) of the AE and statistical CSI of the PE are known, the communication and sensing beamforming problem is formulated with the objective of minimizing the weighted sum of the beampattern matching mean squared error (MSE) and cross-correlation, subject to the secure transmission constraint. To tackle the non-convexity, we propose a semi-definite relaxation (SDR) algorithm and a reduced-complexity zero-forcing (ZF) algorithm. Then, the scenarios are further extended to more general cases with imperfect AE CSI and unknown PE CSI. To further improve the communication performance, the second-stage problem is developed to optimize the secrecy rate threshold under the radar performance constraint. Finally, numerical results demonstrate the superiority of the proposed scheme in terms of sensing and secure communication.
\end{abstract}

\begin{IEEEkeywords}
Integrated sensing and communication (ISAC), physical layer security (PLS), colluding eavesdropping, secrecy rate, beamforming.
\end{IEEEkeywords}

\section{Introduction}

 In recent years, integrated sensing and communication (ISAC) has been widely recognized as an enabling technology for the next-generation wireless networks \cite{1,2}, for applications such as autopilot, unmanned aerial vehicles, and industrial automation \cite{3}. By leveraging shared spectrum, waveforms, and hardware resources, the ISAC can considerably improve the spectral and energy efficiencies. Nevertheless, the repeated use of signals with information embedded for radar sensing may lead to information leaks and potential security challenges in ISAC \cite{5}. The technology of physical layer security (PLS) can be adopted to tackle the eavesdropping security problem \cite{6,7,711,722}. Built upon the theoretical foundation of information theory \cite{8}, PLS considers the physical characteristics of wireless channels, including time-varying characteristics, randomness and reciprocity \cite{9}. The PLS elevates the signal-to-interference-plus-noise ratio (SINR) of target users whilst suppressing the SINR of eavesdroppers (Eves) through techniques such as beamforming and artificial noise (AN) \cite{10,R1,11}.

In contrast to communication-only scenarios, the detected targets may play the role of Eves in secure ISAC systems, leading to more challenging design \cite{12}. In a multi-antenna system, secure beamforming technology has been proposed for increasing the difference in the received signal strength between the legitimate users (LUs) and Eves \cite{13,14,15,16}. In addition to sending communication signals, the transmitter can also send AN to confuse Eves and prevent confidential information leaks. The authors introduced the AN at the transmitter and designed the transmit precoding matrix to minimize the signal-to-noise ratio (SNR) of the Eve under the SINR constraint of LUs \cite{17}. For the problem where the degrees of freedom (DoF) for MIMO radar waveforms is limited by the number of LUs, \cite{18} and \cite{19} developed separate communication and radar waveform precoding matrices in scenarios with multiple Eves. Also, passive reconfigurable intelligent surface (RIS) was introduced in \cite{20} to assist communication in the ISAC system, and the SINR of the radar echo signal was optimized by adjusting the phase shift matrix and beamformer under the SINR thresholds constraints of the LUs and Eves. An active RIS was introduced in \cite{21} to perform secondary empowerment for signal transmission. In addition, the symbol-level precoding was applied in \cite{24} to enhance the signal strength at LUs, while destructive interference is utilized to degrade the eavesdropping signal strength at Eves.

Most of the previous works on PLS mainly have focused on passive eavesdroppers (PEs), which silently eavesdrop on information. When multiple PEs exist, they may individually decode the confidential information or work jointly as colluding eavesdroppers \cite{25,26}. The colluding eavesdropping case would be more practical \cite{27,28} and has not been considered in related ISAC works. In addition, the active eavesdroppers (AEs), who attempt to eavesdrop information and deteriorate the reception of LUs, can be considered in communication scenarios \cite{29,30}. These AEs can destroy communication posing a greater threat to a system than just PEs \cite{31,32,33}. Therefore, in hybrid-colluding scenarios where AEs cooperate with PEs to eavesdrop on confidential information, research on the joint design of information and sensing beamformers is crucial for secure ISAC systems.

Motivated by the works above, in this paper, we propose a joint beamforming design scheme for the ISAC systems in the presence of hybrid-colluding Eves, where one of the sensing targets as AE colludes with a PE. Realistically, the AE will send interference signals to LUs, which lead to AE being discovered by the base station (BS). The PE hides its position and eavesdrops the confidential information without emitting signals. Therefore, it is hard for the BS to acquire the instantaneous channel state information (CSI). As a result, we take several scenarios into account in our work, including perfect and imperfect CSI of the AE, statistical and unknown CSI of the PE, and the efficient secure designs on the hybrid beamformers within these contexts are necessary. The main contributions are listed as follows:
\begin{itemize}
\item In a more practical secure ISAC system, the BS transmits a signal consisting of information symbols and sensing waveform, where the sensing waveform can be also used as artificial noise to interfere with Eves, which enables the DoF for MIMO radar waveform to not be limited by the number of LUs. Under this setup, an alternating optimization-based two stage scheme (AO-TSS) is developed for improving the secure communication performance.
\item In the first stage, the information and sensing beamforming design based on the assumptions of the perfect AE CSI and statistical PE CSI is proposed by minimizing the weighted sum of the beampattern matching mean squared error (MSE) and cross-correlation. This problem is formulated subject to the communication quality of service (QoS) requirement and the PLS constraint. To tackle the non-convexity, we conceive the semi-definite relaxation (SDR) algorithm and low-complexity zero-forcing (ZF) algorithm. Moreover, considering more realistic scenarios, we further design the beamformers under the case of imperfect AE CSI and unknown PE CSI.

\item In the second stage, to improve the secure performance under the optimal radar performance, we further maximize the secrecy rate threshold by performing optimization on SINR thresholds under fixed beamformers. Such a non-convex problem is then solved by the successive convex approximation (SCA) algorithm.
\item The convergence of the proposed AO-TSS can be guaranteed, and the analysis is conducted. Then, simulation results are provided to verify the validity of the proposed AO-TSS. It is shown that employing the proposed algorithms for designing sensing and information beamformers maximizes sensing performance while ensuring secure communication requirements. Additionally, we confirm that the AO-TSS offers advantages over the scheme without two stage.
\end{itemize}

The remainder of this paper is organized as follows: Section \ref{2jie} gives the system model, and Section \ref{3jie} introduces the performance metrics of sensing and secure communication. Section \ref{4jie} formulates a two-stage problem for the beamformers and SINR thresholds designs. Simulation results are presented in Section \ref{5jie}. Finally, we provide the conclusion in Section \ref{6jie}.

{\it{Notations}}: Throughout this paper, vectors and matrices are denoted by bold lowercase and uppercase letters, respectively. ${\mathbb{C}^{M \times K}}$ denotes the space of $M \times K$ matrices with complex entries. $\mathbb{S}_M^ + $ defines the space of $M \times M$ positive semi-definite matrices. ${\bf{I}_{\it{M}}}$ represents an $M \times M$ identity matrix. For a square matrix {\bf{A}}, ${\rm{tr}}({\bf{A}})$ indicates its trace and ${\bf{A}} \succeq 0$ means that {\bf{A}} is positive semi-definite. For a complex matrix {\bf{B}}, ${\rm{rank}}({\bf{B}})$, ${{\bf{B}}^{\rm{T}}}$, ${{\bf{B}}^{\rm{H}}}$ equal its rank, transpose, conjugate transpose, respectively. $\mathbb{E}( \cdot )$ stands for the stochastic expectation, and $\mathcal{CN}({\bf{x}},{\bf{Y}})$ determines the circularly symmetric complex Gaussian (CSCG) random distribution with mean vector {\bf{x}} and covariance matrix {\bf{Y}}.
\section{System Model}\label{2jie}    

Fig. 1 presents a secure ISAC system, which consists of a  half-duplex BS equipped with {\it{M}} antennas in a fully digital array, {\it{K}} single-antenna LUs and {\it{Q}} sensing targets in the presence of colluding Eves. The BS communicates with users and detects targets simultaneously. We consider Eves collusion, where a full-duplex AE and a PE work jointly to decode the confidential information. More specifically, the AE as a target eavesdrops the information while actively emitting jamming signals to compromise the links between the BS and the LUs, and the PE just eavesdrops the confidential information in the system. Let $\mathbb{K}=\{1,...,K\}$ denote the set of LUs, and let $\mathbb{Q}=\{1,...,Q\}$ denote the set of sensing targets, the $Q$-th target is assumed to be the AE.
\begin{figure}[htbp]
\centering
\includegraphics[width=3.4in]{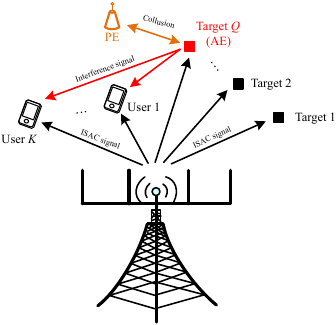}
\caption{The ISAC system model of the secure beamforming with hybrid-colluding eavesdroppers.}
\label{fig_1}
\end{figure}

The discrete-time transmitted signal\footnote{The transmitted signal is composed of communication symbols and sensing signal. For the communication function, the communication signal carries information, while the sensing signal serves as artificial noise to enhance secure communication performance by confusing eavesdroppers. For the radar function, ${\bf{x}}(n)$ is utilized for target detection.} at time slot {\it{n}} is given by \cite{liuxiang}
\begin{align}
\label{1}
{\bf{x}}(n) = {{\bf{W}}_{\rm{s}}}{\bf{s}}(n) + {{\bf{W}}_{\rm{c}}}{\bf{c}}(n),~{\rm{for}}~ n = 1,...,N,
\end{align}
where ${{\bf{W}}_{\rm{s}}} \in {\mathbb{C}^{M \times M}}$ is the sensing beamforming matrix, ${\bf{s}}\left( n \right) = {\left[ {{s_1}\left( n \right), \ldots ,{s_M}\left( n \right)} \right]^{\rm{T}}}$ represents the sensing signal, with ${\rm{\mathbb{E}\{ }}{\bf{s}}(n){\bf{s}}{^{\rm{H}}(n)}{\rm{\} }} = {{\bf{I}}_M}$, and ${{\bf{W}}_{\rm{c}}} = \left[ {{{\bf{w}}_1},{{\bf{w}}_2},{\rm{ }} \ldots ,{{\bf{w}}_K}} \right] \in {\mathbb{C}^{M \times K}}$ indicates the beamforming matrix for message ${\bf{c}}\left( n \right) = {\left[ {{c_1}\left( n \right), \ldots ,{c_K}\left( n \right)} \right]^{\rm{T}}}$, with ${\rm{\mathbb{E}\{ }}{\bf{c}}(n){\bf{c}}{^{\rm{H}}(n)}{\rm{\} }} = {{\bf{I}}_K}$. $N$ is the total number of symbols. We assume that ${\bf{s}}\left( n \right) $ and ${\bf{c}}\left( n \right)$ are statistically independent, and abbreviate ${\bf{s}}\left( n \right)$ and ${\bf{c}}\left( n \right)$ as ${\bf{s}}$ and ${\bf{c}}$, respectively. Then, the covariance matrix of the transmitted signal can be obtained as
\begin{align}
\label{deqn_ex1}
{\bf{R}} = {\rm{\mathbb{E}\{ }}{\bf{x}}(n){\bf{x}}{^{\rm{H}}(n)}{\rm{\} }} ={{\bf{W}}_{\rm{c}}}{\bf{W}}_{\rm{c}}^{\rm{H}}+{{\bf{W}}_{\rm{s}}}{\bf{W}}_{\rm{s}}^{\rm{H}} = {{\bf{R}}_{\rm{c}}} + {{\bf{R}}_{\rm{s}}},
\end{align}
where ${{\bf{R}}_{\rm{c}}} = \sum\nolimits_{k = 1}^K {{{\bf{R}}_k}} $ with ${{\bf{R}}_k} = {{\bf{w}}_k}{\bf{w}}_k^{\rm{H}}$ being the covariance matrix of the $k$-th LU, and ${{\bf{R}}_{\rm{s}}} = \sum\nolimits_{i = K + 1}^{K + M} {{{\bf{R}}_i}} $ is the covariance matrix of the sensing waveform.

Then, the received signal ${\it{y}}_k$ at the $k$-th LU can be expressed as
\begin{align}
\label{3}
{y_k} = {\bf{h}}_k^{\rm{H}}{{\bf{W}}_{\rm{c}}}{\bf{c}} + {\bf{h}}_k^{\rm{H}}{{\bf{W}}_{\rm{s}}}{\bf{s}} + \sqrt {{P_{\rm{a}}}} {{\it{h}}_{{\rm{a}},{\it{k}}}}e + {n_k},
\end{align}
where ${{\bf{h}}_k} \in {\mathbb{C}^{M \times 1}}$ denotes the channel vector from a BS to the $k$-th LU. In addition, we assume that the jamming signals emitted by AE are noise signals, ${P_{\rm{a}}}$ equals the transmit jamming power of the AE, ${{\it{h}}_{{\rm{a}},k}}$ defines the channel from the AE to the $k$-th LU, {\it{e}} means the jamming signal emitted from the AE with ${\rm{E}}[|e{|^2}] = 1$, and $n_k$ is the complex additive white Gaussian noise (AWGN) with zero mean and variance $\sigma _{\rm{c}}^{\rm{2}}$.

For the sensing operation, we consider the Saleh-Valenzuela geometric model \cite{S_V}, which can accurately capture the geometric features in sparse channels. {Then, the signal received at the {\it{Q}}-th target (i.e., AE) is
\begin{align}
\label{4}
{y_{\it{Q}}} = \beta_{\it{Q}} {{\bf{a}}^{\rm{H}}}(\theta_{\it{Q}} ){{\bf{W}}_{\rm{c}}}{\bf{c}} + \beta_{\it{Q}} {{\bf{a}}^{\rm{H}}}(\theta_{\it{Q}} ){{\bf{W}}_{\rm{s}}}{\bf{s}} + {n_{\rm{s}}},
\end{align}
where $\beta_{\it{Q}} $ represents the path loss factor, ${\bf{a}}(\theta_{\it{Q}} )$ indicates the steering vector of the transmit antenna array, and ${n_{\rm{s}}}$ denotes the complex AWGN with zero mean and variance $\sigma _{\rm{s}}^{\rm{2}}$. More specifically, for the AE, we consider the worst case that the self-interference can be eliminated perfectly \cite{self-interference}. Thus the interference will not affect the received signal $y_Q$ at the AE.

Since the PE keeps silent to hide its existence, it is hard for the BS to acquire the instantaneous CSI. Hence, for practical implementation, we assume that only the statistical CSI of the PE is available, which is a generic assumption and has been adopted in most of the works in PLS \cite{PE_CSI1, PE_CSI2, PE_CSI3, PE_CSI4, PE_CSI5, PE_CSI6}. Moreover, we assume that the jamming signal from AE can be eliminated at PE through cooperation relationship \cite{PE_interference}. Then, the signal received at the PE is given by
\begin{align}
\label{5}
{y_{\rm{p}}} = {{\bf{h}}_{\rm{p}}^{\rm{H}}}{{\bf{W}}_{\rm{c}}}{\bf{c}} + {{\bf{h}}_{\rm{p}}^{\rm{H}}}{{\bf{W}}_{\rm{s}}}{\bf{s}} + {n_{\rm{p}}},
\end{align}
where ${{\bf{h}}_{\rm{p}}} \in {\mathbb{C}^{M \times 1}}$ stands for the channel vector from the BS to the PE, and ${n_{\rm{p}}}$ equals the complex AWGN at the PE with zero mean and variance $\sigma _{\rm{p}}^{\rm{2}}$.
\section{Performance Metrics of sensing and Communication}
\label{3jie}
In this section, we introduce the sensing performance metric and give a review of the secrecy rate as the communication performance metric for the considered system.
\subsection{Sensing Performance Metric}                                      
As the sensing performance metric, we consider the weighted sum of the beampattern MSE and cross-correlation. The beampattern MSE allows us to concentrate the signal power and maintain a low sidelobe level \cite{MSE}, and the cross-correlation is adopted to enhance the resolution capability of the radar system for multiple targets \cite{CORR}. First, the transmit beampattern at the angle $\theta$ is defined as
\begin{align}
\label{6}
P(\theta ;{\bf{R}}){\rm{ = }}{{\bf{a}}^{\rm{H}}}(\theta ){\bf{Ra}}(\theta ).
\end{align}
Then, the beampattern MSE which measures the difference between the actual transmit beampattern ${P({\theta _l};{\bf{R}})}$ in the angular domain and the ideal beampattern gain ${\Phi ({\theta _l})}$  at ${\theta _l}$ is defined by
\begin{align}
\label{7}
{L_{\rm{b}}}({\bf{R}},{\delta _1}) = \frac{1}{L}\sum\limits_{l = 1}^L {{{\left| {{ \delta _1}\Phi ({\bar \theta _l}) - P({\bar \theta _l};{\bf{R}})} \right|}^2}},
\end{align}
where ${\delta _1}$ is a scaling factor to be optimized, and $\{ {\bar \theta _l}\} _{l = 1}^L$ denotes the $L$ sampled angle grids covering the detection angular range in [${\rm{-\pi }}$/2, ${\rm{\pi }}$/2].

Furthermore, the cross-correlation between the transmitted signal at any two target directions ${\it{\theta}}_1$ and ${\it{\theta}}_2$ is expected to be small, such that the radar system can perform adaptive localization, i.e.,
\begin{align}
\label{8}
{P_{\rm{c}}}({\theta _1},{\theta _2};{\bf{R}}){\rm{ = }}{{\bf{a}}^{\rm{H}}}({\theta _1}){\bf{Ra}}({\theta _2}).
\end{align}
Then, to improve the resolution capability of the radar system, the mean-squared cross-correlation is computed as
\begin{align}
\label{9}
\begin{array}{*{20}{l}}
{{L_{\rm{c}}}({\bf{R}})}{ = \dfrac{2}{{{Q^2} - Q}}\sum\limits_{p = 1}^{Q - 1} {\sum\limits_{q = p + 1}^Q {{{\left| {{{\bf{a}}^{\rm{H}}}({{\theta }_p}){\bf{Ra}}({{\theta }_q})} \right|}^2}} } ,}
\end{array}
\end{align}
where ${\Omega _{\rm{T}}}=\left\{ {{{\theta }_q}} \right\}_{q = 1}^Q$ represent the given directions of the targets.

Consequently, similar to the approach in \cite{sumloss}, as the sensing performance metric, we adopt the weighted sum of the beampattern MSE and cross-correlation as
\begin{align}
\label{10}
L({\bf{R}},{\delta _1}) = {L_{\rm{c}}}({\bf{R}}) + {\delta _2}{L_{\rm{b}}}({\bf{R}},{\delta _1}),
\end{align}
where $\delta_{2} \in [0,1]$ is the weighting factor. As $\delta_{2}$ is close to 1, the system mainly focuses on the optimization of the quality of beampattern matching. On the contrary, for a small $\delta_{2}$, the system tends to optimize the cross-correlation. In practice, we can determine the specific value of $\delta_{2}$ according to the system requirements. From (\ref{10}), the sensing performance objective is closely related to the covariance matrix ${\bf{R}}$ of the transmitted signal.\\
\subsection{Communication Performance Metrics}                  
For the LUs, according to (\ref{3}), the average SINR of the {\it{k}}-th LU $(k \in \mathbb{K}$) is written as
\begin{align}
{{\gamma _{{\rm{u}},k}} = \frac{{{\bf{h}}_{k}^{\rm{H}}{{\bf{R}}_k}{{\bf{h}}_{k}}}}{{{\bf{h}}_{k}^{\rm{H}}\left( {{\bf{R}} - {{\bf{R}}_k}} \right){{\bf{h}}_{k}} + {P_{\rm{a}}}|{\it{h}}_{{\rm{a}},{\it{k}}}^{}{|^2} + \sigma _{\rm{c}}^{\rm{2}}}},}
\end{align}
where ${\bf{h}}_k^{\rm{H}}\left( {{\bf{R}} - {{\bf{R}}_k}} \right){{\bf{h}}_k}$ represents the radar and multi-user interference (MUI). Then, the corresponding achievable rate of the ${\it{k}}$-th LU can be expressed as
\begin{align}
\label{12}
{R_{{\rm{u}},k}} = {\log _2}(1 + {\gamma _{{\rm{u,}}k}}).
\end{align}

 For the AE and PE, we assume that all the information for the $K$ LUs is the desired. Then, according to (4), the received SINR of the AE is given by \cite{AE_SINR1,AE_SINR2}
\begin{align}
\label{13}
{{\gamma _{\rm{a}}} = \dfrac{{|\beta _{\it{Q}} {|^2}{\bf{a}}{{^{\rm{H}}(\theta _{\it{Q}} )}} {{{\bf{R}}}}_{\rm{c}} {\bf{a}}(\theta _{\it{Q}} )}}{{|\beta_{\it{Q}} {|^2}{\bf{a}}{{^{\rm{H}}(\theta _{\it{Q}} )}} {{{\bf{R}}}}_{\rm{s}} {\bf{a}}(\theta _{\it{Q}} ) + {\sigma _{\rm{s}}^2}}}}.
\end{align}

In practice, since the AE is one of targets in the ISAC system, we assume that the BS can acquire perfect CSI of the AE. However, for the PE, we only know its statistical CSI. Then, from (5),  the received SINR of the PE is obtained by
\begin{align}
\label{14}
\begin{array}{*{20}{l}}
{{\gamma _{\rm{p}}}{ = \dfrac{{ {{{\bf{h}}_{\rm{p}}^{\rm{H}}}{{\bf{R}}}_{\rm{c}}{{\bf{h}}_{\rm{p}}}} }}{{{{{\bf{h}}_{\rm{p}}^{\rm{H}}}{{\bf{R}}}_{\rm{s}}{{\bf{h}}_{\rm{p}}}}  + {\sigma _{\rm{p}}^2}}}}}.
\end{array}
\end{align}
Further, when the single-antenna AE and PE cooperate, they can be seen as one Eve with multiple antennas, which could enhance the eavesdropping capability. By performing maximal ratio combining, the SINR of AE and PE can be summed up \cite{MRC}. Then, the achievable rate of Eve is expressed as
\begin{align}
\label{16}
{R_{\rm{e}}} = {\log _2}(1 + {\gamma _{\rm{a}}} + {\gamma _{\rm{p}}}).
\end{align}
Therefore, the secrecy rate is defined as \cite{SR}
\begin{align}
\label{17}
{C_{\rm{s}}} = \mathop {\min }\limits_{k \in \mathbb{K}} {({R_{{\rm{u}},k}} - {R_{\rm{e}}})^ + },
\end{align}
where ${(x)^ + } = \max \{ 0,x\}$. As a fairness performance metric, the secrecy rate ${C_{\rm{s}}}$ guarantees that each LU can obtain satisfactory security \cite{fairness}, and quantifies the maximum achievable data rate for reliably transmitting confidential messages from the BS to LUs, ensuring that Eves cannot intercept confidential messages, even if they possess infinite computational power to intercept the signals.
\section{the beamforming design for \\ hybrid-colluding eavesdroppers}        
\label{4jie}
In this section, we propose the AO-TSS to maximize the sensing and communication performance. By obtaining the information and sensing beamforming matrices with a given communication SINR thresholds, the first-stage problem based on the assumptions of the perfect AE CSI and statistical PE CSI is to minimize the difference between the desired transmit beampattern and the actual beampattern. Subsequently, we design the secure ISAC waveforms under scenarios where the AE CSI is imperfectly known and the PE CSI is unknown. Then, under the condition of the known transmit beamforming matrices, the second-stage problem maximizes the secrecy rate by adjusting the communication SINR thresholds of LUs, AE and PE. The optimization problems of these two stages are optimized alternately until convergence.
\subsection{Transmit Beamforming Design with perfect AE CSI and statistical PE CSI}            
We jointly optimize the transmit information covariance matrix ${{{\bf{R}}_k}}$ ($k \in \mathbb{K}$) and the sensing covariance matrix ${{{\bf{R}}_{\rm{s}}}}$, with the objective of minimizing the weighted sum of the beampattern MSE and cross-correlation, subject to the communication QoS requirement of each LU, PLS level constraints as well as the transmit power budget constraint. Accordingly, the optimization problem can be formulated as
\begin{align}
\label{18}
{\bf{P1:}}{\mathop {\min }\limits_{\{ {{\bf{R}}_k}\}, {\bf{R}},{\delta _1}} }&{L({\bf{R}},{\delta _1})}\tag{17a}\\
\hspace{-5cm}{{\rm{s}}.{\rm{t}}.}~~~&{{\gamma _{{\rm{u}},k}} \ge {\varepsilon _{{\rm{u}},k}},\forall k},\tag{17b}\\
{}&{{\gamma _{\rm{a}}} \le {\varepsilon _{\rm{a}}}},\tag{17c}\\
{}&{{\rm{Pr}}\left( {{\gamma _{\rm{p}}} \le {\varepsilon _{\rm{p}}}} \right) \ge \tau },\tag{17d}\\
{}&{{\rm{tr}}({\bf{R}}){\rm{ = }}{P_0}},\tag{17e}\\
{}&{{{\bf{R}}_k} \in \mathbb{S}_M^ + ,{\rm{rank}}({{\bf{R}}_k}) = 1,\forall k},\tag{17f}\\
{}&{{\bf{R}}_{\rm{s}}  \in \mathbb{S}_M^ + ,{\delta _1} \ge {\rm{0}}},\tag{17g}
\end{align}
where the constraint (17b) ensures the reliability of communication services for LUs, ${\varepsilon _{\rm{u,{\it{k}}}}}$ represents the SINR threshold of the {\it{k}}-th LU, and ${\it{P}}_0$ is the power budget. For secure communication, the constraint (17c) requires that the received SINR of the AE should be less than the threshold ${\varepsilon _{\rm{a}}}$. The constraint (17d) sets the minimum outage requirement for the PE. Specifically, the received SINR of the PE is required to be smaller than the threshold ${{\varepsilon _{\rm{p}}}}$ with at least probability $\tau$. (17f) and (17g) require that the covariance matrices of each LU and sensing waveform are positive semi-definite, and the rank of the covariance matrix for each LU is 1.\\
${\it{Remark}}$ 1. It is worth noting that the proposed design can be applied to more scenarios, including those with only PE or only AE. Specifically, in this scenario, without AE, we just need to remove the SINR constraint (17c) related to the AE in P1. Additionally, without PE, we just need to remove the minimum outage requirement (17d) for the PE in P1.

Here, the probabilistic constraint (17d) involves the non-convex optimization variables. To make the problem tractable, we convert (17d) to a convex constraint according to the following Lemma.

${\it{Lemma~1}}$: Let us assume that the channel of the PE is modeled as independent and identical distributed (i.i.d.) Rayleigh variables as ${{\bf{h}}_{\rm{p}}} = {[{h_{{\rm{p}},1}},...,{h_{{\rm{p}},M}}]^{\rm{T}}}\sim \mathcal{CN}({\bf{0}},c{{\bf{I}}_M})$. Then the probabilistic constraint (17d) is conservatively transformed as
\begin{align}
\label{19}
\setcounter{equation}{17}
\begin{array}{*{20}{l}}
{ {\lambda _{\max }}\left( {{{\bf{R}}_{\rm{c}}} - {\varepsilon _{\rm{p}}}{{\bf{R}}_{\rm{s}}}} \right) \le \Phi _M^{ - 1}\left( {1 - \tau } \right)\dfrac{{{\varepsilon _{\rm{p}}}\sigma _{\rm{p}}^2}}{c},}
\end{array}
\end{align}
where $\Phi _M^{ - 1}\left(  \cdot  \right)$ denotes the inverse cumulative distribution function (CDF) of an inverse central chi-square random variable with 2{\it{M}} DoF.

{\it{Proof}}: Please refer to Appendix A.

For the constraints (17b)$-$(17d), which involve intricate fractional operations, we employ straightforward mathematical transformations to simplify these three constraints as
\begin{align}
\label{20}
\left( {1 + \frac{1}{{{\varepsilon _{{\rm{u}},k}}}}} \right){\bf{h}}_k^{\rm{H}}{{\bf{R}}_k}{{\bf{h}}_k} \ge {\bf{h}}_k^{\rm{H}}{\bf{R}}{{\bf{h}}_k} + {P_{\rm{a}}}|h_{{\rm{a}},k}^{}{|^2} + \sigma _{\rm{c}}^{\rm{2}},\forall k,
\end{align}
\begin{align}
\label{21}
\left( {1 + \frac{1}{{{\varepsilon _{\rm{a}}}}}} \right){{\bf{a}}^{\rm{H}}}({\theta _Q}){{\bf{R}}_{\rm{c}}}{\bf{a}}({\theta _Q}) \le {{\bf{a}}^{\rm{H}}}({\theta _Q}){\bf{Ra}}({\theta _Q}) + \frac{{\sigma _{\rm{s}}^{\rm{2}}}}{{{{\left| {{\beta _Q}} \right|}^2}}},
\end{align}
\begin{align}
\label{22}
{\lambda _{\max }}\left( {\left( {1 + {\varepsilon _{\rm{p}}}} \right) {{{\bf{R}}}}_{\rm{c}} - {\varepsilon _{\rm{p}}}{\bf{R}}} \right) \le \Phi _M^{ - 1}\left( {1 - \tau } \right){\varepsilon _{\rm{p}}}{\sigma ^2_{\rm{p}}}.
\end{align}
It can be observed that (19)$-$(21) are convex affine constraints. Accordingly, the optimization problem is reformulated as
\begin{align}
\label{23}
\begin{array}{*{20}{l}}
{\bf{P2:}}{\mathop {\min }\limits_{\{ {{\bf{R}}_k}\}, {\bf{R}},{\delta _1}} }&{L({\bf{R}},{\delta _1})}\\
\hspace{14mm}{{\rm{s}}.{\rm{t}}.}&{(19)-(21)},\\
{}&{{\rm{tr}}({\bf{R}}){\rm{ = }}{P_0}},\\
{}&{{{\bf{R}}_k} \in \mathbb{S}_M^ + ,{\rm{rank}}({{\bf{R}}_k}) = 1,\forall k },\\
{}&{{\bf{R}}_{\rm{s}} \in \mathbb{S}_M^ + },\\
{}&{{\delta _1} \ge {\rm{0}}}.
\end{array}
\end{align}

Still problem P2 is non-convex. To solve this, we present the following two algorithms.
\subsubsection{SDR algorithm}

we initially obtain the optimal solution $\hat \Xi  = \{ {\bf{\hat R}}_k\}$ without considering the rank-1 constraint. Then we restore the solution $\tilde \Xi  = \{ {\bf{\tilde R}}_k\} $ that satisfies the rank-1 condition from $\hat \Xi$. Neglecting the rank-1 constraint in P2, we formulate a convex relaxation problem as
\begin{align}
\label{233}
\begin{array}{*{20}{l}}
{\bf{P3:}}{\mathop {\min }\limits_{\{ {{\bf{R}}_k}\}, {\bf{R}},{\delta _1}} }&{L({\bf{R}},{\delta _1})}\\
\hspace{14mm}{{\rm{s}}.{\rm{t}}.}&{(19)-(21)},\\
{}&{{\rm{tr}}({\bf{R}}){\rm{ = }}{P_0}},\\
{}&{{{\bf{R}}_k} \in \mathbb{S}_M^ + ,\forall k},\\
{}&{{\bf{R}}_{\rm{s}}  \in \mathbb{S}_M^ + },\\
{}&{{\delta _1} \ge {\rm{0}}}.
\end{array}
\end{align}
Here P3 is a standard quadratic semi-definite program (QSDP) problem, and the optimal solution $\hat \Xi$ can be obtained by a standard convex optimization toolbox such as CVX\cite{cvx}. If ${\bf{\hat R}}_k$ is not rank-1, the following proposition addresses that the optimal solutions ${\bf{\tilde R}}_k$ to P2 can be calculated and approximated.

{\bf{Proposition 1}}: Based on the obtained optimal solution $\hat \Xi$ to the QSDP problem P3, we can always construct the optimal solution $\tilde \Xi$ with rank 1 to the original non-convex problem P2 as
\begin{align}
\label{24}
{{{\bf{\tilde w}}}_k} = {\left( {{\bf{h}}_k^{\rm{H}}{{{\bf{\hat R}}}_k}{{\bf{h}}_k}} \right)^{ - 1/2}}{{{\bf{\hat R}}}_k}{{\bf{h}}_k},
\end{align}
and
\begin{align}
\label{244}
{{{\bf{\tilde R}}}_k} = {{{\bf{\tilde w}}}_k}{\bf{\tilde w}}_k^{\rm{H}}.
\end{align}
{\it{Proof}}: Please refer to Appendix B.

Proposition 1 validates the solution $ {{{\bf{\tilde R}}}}$ and ${{\left\{ {{\bf{\tilde w}}_k} \right\}}_{k \in \mathbb{K}}} $ to P2 obtained by the SDR algorithm are tight. Therefore, we can obtain the global optimum for ISAC secure transmission by the proposed SDR algorithm.
\subsubsection{Low complexity ZF algorithm}   
The computational burden to solve P3 with the SDR algorithm arises from the presence of multiple semi-definite constraints. Furthermore, the number of semi-definite constraints increases with LUs. Taking this into consideration, we propose a reduced-complexity ZF algorithm. The ZF constraints encompass both communication and radar. Under the condition of zero communication MUI, ${\bf{W}}_{\rm{c}}$ should satisfy
\begin{align}
\label{25}
{{\bf{H}}_{\rm{u}}}{{\bf{W}}_{\rm{c}}} = {\rm{diag}}(\sqrt {{\rho _1}} ,...,\sqrt {{\rho _K}} ),
\end{align}
where ${{\bf{H}}_{\rm{u}}} = {\left[ {{{\bf{h}}_1},...,{{\bf{h}}_K}} \right]^{\rm{H}}} \in {\mathbb{C}^{K \times M}}$ denotes the channel matrix from the BS to LUs, ${\rho _k}$ represents the required transmit power for sending information to the {\it{k}}-th LU. Similarly, under the condition of zero radar interference, the sensing beamforming matrix ${\bf{W}}_{\rm{s}}$ should meet
\begin{align}
\label{26}
{{\bf{H}}_{\rm{u}}}{{\bf{W}}_{\rm{s}}} = {{\bf{0}}_{K \times M}}.
\end{align}
Accordingly, we have
\begin{align}
\label{27}
{{\bf{H}}_{\rm{u}}}{\bf{RH}}_{\rm{u}}^{\rm{H}} = {{\bf{H}}_{\rm{u}}}{{\bf{R}}_{\rm{c}}}{\bf{H}}_{\rm{u}}^{\rm{H}} + {{\bf{H}}_{\rm{u}}}{{\bf{R}}_{\rm{s}}}{\bf{H}}_{\rm{u}}^{\rm{H}} = {\rm{diag}}\left( {{\rho _1},...,{\rho _K}} \right),
\end{align}
and
\begin{align}
\label{266}
{{\bf{H}}_{\rm{u}}} {{\bf{R}}_{\rm{s}}}{\bf{H}}_{\rm{u}}^{\rm{H}} = {{\bf{H}}_{\rm{u}}}{{\bf{W}}_{\rm{s}}}{\bf{W}}_{\rm{s}}^{\rm{H}}{\bf{H}}_{\rm{u}}^{\rm{H}} = {{\bf{0}}}.
\end{align}

After imposing the ZF constraints, the individual matrix variable ${{\bf{R}}_k}$ can be removed from the SINR constraints of LUs. Then, (\ref{20}) can be reformulated as
\begin{align}
\label{28}
{\rho _k} \ge {\varepsilon _{\rm{u,{\it{k}}}}}\left( {{P_{\rm{a}}}|{{\it{h}}_{{\rm{a}},k}}{|^2} + {\sigma _{\rm{c}}^2}} \right).
\end{align}
It can be seen that the covariance matrix constraint of each LU is converted to the transmit power constraint.
Finally, the problem can be rewritten as
\begin{align}
\label{31}
\begin{array}{*{20}{l}}
{{\bf{P4}}:\mathop {\min }\limits_{{{\bf{R}}_{\rm{c}}},{\bf{R}},{\bf{\rho }},{\delta _1}} }&{L({\bf{R}},{\delta _1})}\\
\hspace{14mm}{{\rm{s}}.{\rm{t}}.}&{(20),(21),(\ref{28})},\\
{}&{{{\bf{H}}_{\rm{u}}}{{\bf{R}}_{\rm{c}}}{\bf{H}}_{\rm{u}}^{\rm{H}} = {\rm{diag}}\left( {{\rho _1},...,{\rho _K}} \right)},\\
{}&{{{\bf{H}}_{\rm{u}}}({\bf{R}} - {{\bf{R}}_{\rm{c}}}){\bf{H}}_{\rm{u}}^{\rm{H}} = {{\bf{0}}}},\\
{}&{{{\bf{R}}_{\rm{c}}} \in \mathbb{S}_M^ + , {{\bf{R}}_{\rm{s}}} \in \mathbb{S}_M^ + },\\
{}&{{\delta _1} \ge {\rm{0}}}.
\end{array}
\end{align}
 Obviously, P4 is a QSDP problem, which can be solved by CVX. Therefore, next we need to recover the beamforming matrix ${{\bf{\tilde W}}_{\rm{c}}}$ from the optimal solutions ${\bf {\tilde R}}$ and ${{\bf{\tilde R}}_{\rm{c}}}$.

 Inspired by \cite{liuxiang}, we propose the following processes to construct the sensing covariance matrix and communication beamforming matrices. First, exploiting the properties of semi-definite matrices, we consider the Cholesky decomposition of ${{\bf{\tilde R}}_{\rm{c}}}$ as
 \begin{align}
\label{311}
 {{\bf{\tilde R}}_{\rm{c}}} = {{\bf{D}}}{\bf{D}}^{\rm{H}},
 \end{align}
 where ${\bf{D}}\in {\mathbb{C}^{M \times M}}$ is a lower triangular matrix. Then, we perform QR decomposition for $ {\bf{H}}_{\rm{u}}{{\bf{D}}}\in {\mathbb{C}^{K \times M}}$ as
\begin{align}
\label{32}
{\bf{H}}_{\rm{u}}{{\bf{D}}}  = \left\{ \begin{array}{l}
{{\bf{U}}_{\rm{1}}}\left[ \begin{array}{l}
{{\bf{B}}_{\rm{U}}}\\
{{\bf{0}}_{\left( {K - M} \right) \times M}}
\end{array} \right],~{\rm{if}}~ K \ge M\\
\left[ {{{\bf{B}}_{\rm{L}}},{{\bf{0}}_{K \times \left( {M - K} \right)}}} \right]{{\bf{U}}_{\rm{2}}},~{\rm{if}}~ K < M
\end{array} \right.,
\end{align}
where ${{\bf{U}}_{\rm{1}}} \in {\mathbb{C}^{K \times K}}$ and ${{\bf{U}}_{\rm{2}}} \in {\mathbb{C}^{M \times M}}$ are unitary matrices, ${{\bf{B}}_{\rm{U}}} \in {\mathbb{C}^{M \times M}}$ represents an upper triangular matrix, and ${{\bf{B}}_{\rm{L}}} \in {\mathbb{C}^{K \times M}}$ indicates a lower triangular matrix. For simplicity, we only consider the case of $K<M$. the case of $K \ge M$ can be obtained by using the same method. Denoting ${{\bf{U}}^{\rm{H}}_2} = \left[ {{\bf{\tilde U}},{\bf{\hat U}}} \right]$, where ${\bf{\tilde U}}$ is the matrix composed of the first $K$ columns of ${{\bf{U}}^{\rm{H}}_2}$, the communication beamforming matrix can be expressed as
\begin{align}
\label{33}
{{{\bf{\tilde W}}}_{\rm{c}}} = {\bf{D\tilde U}}.
\end{align}
Besides, the radar covariance matrix can be obtained as
\begin{align}
\label{34}
{{\bf{\tilde W}}_{\rm{s}}}{{\bf{\tilde W}}_{\rm{s}}^{\rm{H}}} = {\bf{\tilde R}} - {{\bf{\tilde W}}_{\rm{c}}}{{\bf{\tilde W}}_{\rm{c}}^{\rm{H}}}.
\end{align}
Next, we will analyze the feasibility and effectiveness of the proposed beamforming design method by introducing the following proposition.

{\bf{Proposition 2}}: With the optimal solutions ${\bf{\tilde R}}$ and ${{\bf{\tilde R}}_{\rm{c}}}$ of problem P4, ${{\bf{\tilde W}}_{\rm{c}}}$ and ${{\bf{\tilde W}}_{\rm{s}}}$ computed in (34) and (35) are also the optimal beamforming matrices of P4, and satisfy the ZF constraints.\\
{\it{Proof}}: Please refer to Appendix C.

Proposition 2 illustrates that $ {{{\bf{\tilde R}}}}$ is the global optimum to P4, and ${{\bf{\tilde W}}_{\rm{c}}}$ obtained by the ZF algorithm is also globally optimal to P4. The detailed procedures of ZF algorithm are summarized in {\bf Algorithm 1}.
\begin{table}[!htbp]
\begin{tabular}{l}
\toprule[1pt] 
{\bf{Algorithm 1}} Solving the beamforming problem P4 via ZF~~~~~~~~~~~~ \\
\midrule 
1. Based on (\ref{27}) and (\ref{266}), transform (\ref{20}) into (\ref{28}). \\
2. Construct the optimization problem P4.\\
3. Compute the optimal solution of P4 via CVX. \\
4. Compute the Cholesky decomposition of ${{\bf{\tilde R}}_{\rm{c}}}$. \\
5. Compute ${{\bf{\tilde w}}_1},...,{{\bf{\tilde w}}_K}$ by (\ref{33}).\\
\bottomrule[1pt] 
\end{tabular}
\end{table}
 \subsubsection{Complexity analysis}
We analyze the worst-case computational complexity of the SDR and ZF algorithms via the interior-point method \cite{complexity}. The primary computation burden of these two algorithms originates from solving the QSDP problems P3 and P4. Specifically, P3 is a convex problem with $\phi_{1}=(K + 1){M^2} + 1$ optimization variables, $(K + 4)$ affine constraints and $(K + 1)$ linear matrix inequality (LMI) constraints. The number of iterations in the interior-point method equals ${I_1}=O(\sqrt {M(K + 1)} \ln (1/ \varepsilon))$, where $\varepsilon $ represents the accuracy of the tolerable maximum duality gap to guarantee the optimality. Then, the cost per-iteration equals $O(\phi_{1} ((K + 1){M^3} + \phi_{1} (K + 1){M^2} + {\phi_{1} ^2}))$. Finally, the total computational complexity becomes $O(\sqrt {M(K + 1)} \ln (1/ \varepsilon) \phi_{1} ((K + 1){M^3} + \phi_{1} (K + 1){M^2} + {\phi_{1} ^2}))$.

Similarly, the number of optimization variables $\phi_{2} $ in P4 is $2{M^2} + 2$, and P4 involves $2KM + K + 3$ affine constraints and $2$ LMI constraints with size $M$. The number of iterations in the interior-point method equals ${I_2}=O(\sqrt {2M} \ln (1/ \varepsilon))$. Then, the cost per-iteration is $O(\phi_{2} (2{M^3} + 2\phi_{2} {M^2} + {\phi_{2} ^2}))$. Finally, the total computational complexity becomes $O(\sqrt {2M} \ln (1/ \varepsilon) \phi_{2} (2{M^3} + 2\phi_{2} {M^2} + {\phi_{2} ^2}))$.

The computational complexity of these algorithms is listed in Table \ref{complexity_table}. From Table \ref{complexity_table}, we observe that the complexity of the SDR algorithm increases with the number of LUs due to the semi-definite programs associated with communication SINR constraints. In contrast to the SDR, the number of variable elements and LMI constraints is significantly lowers in the ZF algorithm. Moreover, the complexity of the ZF algorithm does not increase with the number of LUs. Therefore, the ZF algorithm may be more suitable for large-scale systems.
\begin{table}[H]
\belowrulesep=0pt
\aboverulesep=0pt
\renewcommand\arraystretch{1.5}
\caption{{computational complexity for the SDR and ZF algorithms}}
\label{complexity_table}
\centering
\begin{tabular}{c|c}
\toprule
${\bf{Algorithm}}$&${\bf{Computational~complexity}}$ \\
\midrule
SDR&$O({I_1}{\phi _1}((K + 1){M^3} + {\phi _1}(K + 1){M^2} + \phi _1^2))$ \\
ZF&$O({I_2}{\phi _2}(2{M^3} + 2{\phi _2}{M^2} + \phi _2^2))$ \\
\bottomrule
\end{tabular}
\end{table}
}
\subsection{Beamforming Design with Imperfect AE CSI and Unknown PE CSI}     
Considering complex real-world issues such as clutter interference and the hidden characteristic of PE, obtaining perfect CSI of AE and statistical CSI of PE may be challenging. Thus, for the AE (the $Q$-th target), we assume that the direction of the $Q$-th target is approximately known by the BS within an angular interval of $\left[ {{\theta _Q} - \Delta ,{\theta _Q} + \Delta } \right]$ \cite{CSI_DOA}, where $\Delta$ represents the associated angle uncertainty. Furthermore, for the PE, we consider a scenario where  CSI of the PE is unknown. In this scenario, we aim to design a relaxed beamforming scheme to ensure secure transmission.
\subsubsection{Design with uncertainty in the direction of AE}
The direction uncertainty of the AE affects both the sensing performance and the PLS constraint of AE in P2. To improve the sensing performance, taking into account angular uncertainty, the BS should form wide main-lobe to encompass all potential directions of the AE by broadening the beamwidth of the ideal beampattern ${\Phi ({\theta _l})}$ in (\ref{7}). Moreover, for the secure communication, since the AE may be located in any direction ${\hat \theta _Q}$ within the angular interval, we need to ensure a satisfactory secrecy rate for every potential direction. Consequently, the SINR constraint (17c) should be modified according to
\begin{align}
\label{AE_un}
{\dfrac{{|{\beta _Q} {|^2}{\bf{a}}{{^{\rm{H}}({\hat \theta _Q})}} {{{\bf{R}}}}_{\rm{c}} {\bf{a}}({\hat \theta _Q})}}{{|{\beta _Q} {|^2}{\bf{a}}{{^{\rm{H}}({\hat \theta _Q})}} {{{\bf{R}}}}_{\rm{s}} {\bf{a}}({\hat \theta _Q}) + {\sigma _{\rm{s}}^2}}}}\le {\varepsilon _{\rm{a}}},\forall {\hat \theta _Q} \in \left[ {{\theta _Q} - \Delta ,{\theta _Q} + \Delta } \right].
\end{align}
It can be observed that the angular uncertainty introduces more constraints similar to (17c) over the associated angular interval. Obviously, the SDR and ZF algorithms are also capable of handling the modified constraints (\ref{AE_un}).
\subsubsection{Design with unknown PE CSI}
In practice, the BS may be unaware of the existence of PE, resulting in the inability to acquire its CSI knowledge. In such case, we try to guarantee the secure performance by limiting the SINR of PE in each possible direction ${\hat \theta _{\rm{p}}}$ \cite{multi-angle}. Let ${{\bar \Omega }_{\rm{P}}}$ be the grid sampled in the range of possible PE directions which satisfies ${\bar \Omega _{\rm{P}}} \cap {\Omega _{\rm{T}}} = \emptyset $. Consequently, the secure outage constraint (17d) should be modified as
\begin{align}
\label{PE_un}
\dfrac{{{{\left| {{\beta _{\rm{p}}}} \right|}^2}{{\bf{a}}^{\rm{H}}}({\hat \theta _{\rm{p}}}){{\bf{R}}_{\rm{c}}}{\bf{a}}({\hat \theta _{\rm{p}}})}}{{{{\left| {{\beta _{\rm{p}}}} \right|}^2}{{\bf{a}}^{\rm{H}}}({\hat \theta _{\rm{p}}}){{\bf{R}}_{\rm{s}}}{\bf{a}}({\hat \theta _{\rm{p}}}) + \sigma _{\rm{p}}^{\rm{2}}}} \le {\varepsilon _{\rm{p}}},\forall {\hat \theta _{\rm{p}}} \in {\bar \Omega _{\rm{P}}}.
\end{align}
Similarly, (\ref{PE_un}) can also be handled by the SDR/ZF algorithm.

In terms of the computational complexity, compared to the beamforming design in Subsection VI-A, the designs in the relaxed cases involve adding $N_{{\rm{A}}}$ and $N_{{\rm{P}}}$ LMI constraints with size $M$, respectively, where $N_{{\rm{A}}}$ and $N_{{\rm{P}}}$ represent the number of the AE SINR constraints (\ref{AE_un}) and the PE SINR constraints (\ref{PE_un}), respectively. Moreover, the number of iterations for the SDR algorithm in these cases equals ${I_3}=O(\sqrt {M(K + {N_{\rm{A}}})} \ln (1/ \varepsilon) )$ and ${I_4}=O(\sqrt {M(K + {N_{\rm{P}}}+1)} \ln (1/ \varepsilon))$, respectively. When using the ZF algorithm, the number of iterations is ${I_5}=O(\sqrt {M(1 + {N_{\rm{A}}})} \ln (1/ \varepsilon))$ and ${I_6}=O(\sqrt {M(2 + {N_{\rm{P}}})} \ln (1/ \varepsilon))$. Subsequently, the increased computational complexity is listed in Table \ref{complexity_table2}.
\begin{table}[H]
	\belowrulesep=0pt
	\aboverulesep=0pt
	\renewcommand\arraystretch{1.5}
	\caption{\textbf {additional computational complexity under the relaxed two cases}}
	\label{complexity_table2}
	\centering
	\begin{tabular}{c|c|c}
		\toprule
		&AE&PE \\
		\midrule
		SDR&$O\left( {{\phi _1}{N_{\rm{A}}}{M^2}\left( {{I_3}M + 1} \right)} \right)$&$O\left( {{\phi _1}{N_{\rm{P}}}{M^2}\left( {{I_4}M + 1} \right)} \right)$ \\
		ZF&$O\left( {{\phi _2}{N_{\rm{A}}}{M^2}\left( {{I_5}M + 1} \right)} \right)$&$O\left( {{\phi _2}{N_{\rm{P}}}{M^2}\left( {{I_6}M + 1} \right)} \right)$ \\
		\bottomrule
	\end{tabular}
\end{table}

\subsection{Secrecy Rate Maximization}     
So far, we have derived the optimal covariance matrix ${\bf{\tilde R}}$ and beamforming matrix ${{\bf{\tilde W}}_{\rm{c}}}$ for the transmitted signal under fixed SINR thresholds. It can be observed that the secrecy rate is not maximized. Subsequently, we need to individually adjust each SINR threshold to enhance secure performance under fixed ${\bf{\tilde R}}$ and ${{\bf{\tilde W}}_{\rm{c}}}$. Therefore, according to (16), the optimization objective for the second-stage problem can be formulated as
\begin{align}
\label{35}
\mathop {\max }\limits_{\{ {\varepsilon _{{\rm{u}},k}}\} ,{\varepsilon _{\rm{a}}},{\varepsilon _{\rm{p}}}} \mathop {\min }\limits_{k } {\log _2}(1 + {\varepsilon _{{\rm{u}},k}}) - {\log _2}(1 + {\varepsilon _{\rm{a}}} + {\varepsilon _{\rm{p}}}).
\end{align}
To ensure the secure performance, we add QoS constraint (\ref{20}) and PLS constraints ((\ref{21}), (\ref{22})) to restrict the range of SINR thresholds, where ${\bf{\tilde R}}$ and ${{\bf{\tilde W}}_{\rm{c}}}$ are given. And we construct the second-stage optimization problem as
\begin{align}
\label{36}
&{\bf{P5:}}{\mathop {\max }\limits_{\{ {\varepsilon _{{\rm{u}},k}}\} ,{\varepsilon _{\rm{a}}},{\varepsilon _{\rm{p}}}} \mathop {\min }\limits_k {\log _2}(1 + {\varepsilon _{{\rm{u}},k}}) - {\log _2}(1 + {\varepsilon _{\rm{a}}} + {\varepsilon _{\rm{p}}})}\tag{39a}\\
&~{{\rm{s.t.}}}~{\left( {1 + 1/{\varepsilon _{{\rm{u,}}{k}}}} \right){\bf{h}}_{k}^{\rm{H}}{{{\bf{\tilde R}}}_k}{{\bf{h}}_{k}}\ge {\bf{h}}_{k}^{\rm{H}}{{{\bf{\tilde R}}}}{{\bf{h}}_{k}} + {P_{\rm{a}}}|{\it{h}}_{{\rm{a}},{\it{k}}}^{}{|^2} + \sigma _{\rm{c}}^{\rm{2}},\forall k},\tag{39b}\\
&~~~~~~{\left( {1 + 1/{\varepsilon _{\rm{a}}}} \right){{\bf{a}}^{\rm{H}}}(\theta_{\it{Q}} ){{{\bf{\tilde R}}}_{\rm{c}}}{\bf{a}}(\theta_{\it{Q}} ) \le {{\bf{a}}^{\rm{H}}}(\theta_{\it{Q}} ){\bf{\tilde Ra}}(\theta_{\it{Q}} ) + \dfrac{{\sigma _{\rm{s}}^{\rm{2}}}}{{{{\left| \beta_{\it{Q}}  \right|}^2}}}},\tag{39c}\\
&~~~~~~{{\lambda _{\max }}\left( {\left( {1 + {\varepsilon _{\rm{p}}}} \right){{{\bf{\tilde R}}}_{\rm{c}}} - {\varepsilon _{\rm{p}}}{\bf{\tilde R}}} \right) \le \Phi _M^{ - 1}\left( {1 - \tau } \right){\varepsilon _{\rm{p}}}\sigma _{\rm{p}}^2}.\tag{39d}
\end{align}

The optimization problem is non-convex due to the second term in the objective. By introducing an auxiliary variable ${\varepsilon _{\rm{e}}} = {\varepsilon _{\rm{a}}} + {\varepsilon _{\rm{p}}}$ and exploiting the first-order Taylor expansion at point $\varepsilon _{\rm{e}}^{(r)}$, which is the optimal value of ${\varepsilon _{\rm{e}}}$ at the {\it{r}}-th iteration, we have
\begin{align}
\setcounter{equation}{39}
\label{37}
{\log _2}(1 + {\varepsilon _{\rm{e}}}) \le {\log _2}(1 + \varepsilon _{\rm{e}}^{(r)}) + \frac{1}{{(1 +\varepsilon _{\rm{e}}^{(r)})\ln 2}}({\varepsilon _{\rm{e}}} - \varepsilon _{\rm{e}}^{(r)}).
\end{align}
Thus, P5 can be approximated by
\begin{align}
\label{38}
\begin{array}{*{20}{l}}
{\bf{P6:}}{\mathop {\max }\limits_{\{ {\varepsilon _{{\rm{u}},k}}\} ,{\varepsilon _{\rm{a}}},{\varepsilon _{\rm{p}}},{\varepsilon _{\rm{e}}}} }&{\mathop {\min }\limits_{k} {\log _2}(1 + {\varepsilon _{{\rm{u}},k}}) - \frac{{{\varepsilon _{\rm{e}}}}}{{(1 + \varepsilon _{\rm{e}}^{(r)})\ln 2}}}\\
\hspace{16mm}{{\rm{s}}.{\rm{t}}.}&{(39{\rm{b}})-(39{\rm{d}}),}\\
{}&{{\varepsilon _{\rm{e}}} = {\varepsilon _{\rm{a}}} + {\varepsilon _{\rm{p}}}}.
\end{array}
\end{align}
Then, by introducing a variable $\omega$, P6 is reformulated as
\begin{align}
\label{39}
\begin{array}{*{20}{l}}
{\bf{P7:}}{\mathop {\max }\limits_{\{ {\varepsilon _{{\rm{u}},k}}\} ,{\varepsilon _{\rm{a}}},{\varepsilon _{\rm{p}}},{\varepsilon _{\rm{e}}},\omega } }&\omega \\
\hspace{16mm}{{\rm{s}}.{\rm{t}}.}&{{\log _2}(1 + {\varepsilon _{{\rm{u}},k}}) - \frac{{{\varepsilon _{\rm{e}}}}}{{(1 + \varepsilon _{\rm{e}}^{(r)})\ln 2}} \ge \omega , \forall k,}\\
{}&{(39{\rm{b}})-(39{\rm{d}}),}\\
{}&{{\varepsilon _{\rm{e}}} = {\varepsilon _{\rm{a}}} + {\varepsilon _{\rm{p}}}}.
\end{array}
\end{align}
P7 is now convex and can be efficiently solved by the CVX toolbox.

We can observe that P7 just involves $2K+3$ affine constraints and the total computational complexity equals $O(\sqrt {2K} \ln \frac{1}{\varepsilon } {\phi _3}(2K({\phi _3} + 1) + \phi _3^2))$, where ${\phi _3}=K+4$ represents the number of optimization variables. Obviously, in the optimization process of the two-stage problem, the computational complexity of the second-stage problem solved is much lower than that of the first-stage problem solved by the SDR/ZF algorithm.

The proposed AO-TSS is presented in {\bf Algorithm 2}, where the first and second stage correspond to steps 4-5 and steps 6-12, respectively.
\begin{table}[!htbp]
	\begin{tabular}{l}
		\toprule[1pt] 
		{\bf{Algorithm 2}} {The proposed AO-TSS} \\
		\midrule 
		{\bf{Initialization:}} $\left\{ {\varepsilon _{{\rm{u,}}k}} \right\},\varepsilon _{\rm{a}},\varepsilon _{\rm{p}}, {\iota _1}, {\iota _2}$.\\
		1.~Set ${r_1} \leftarrow 0$. \\
		2.~{\bf{repeat}}\\
		3.~~~~${r_1} \leftarrow {r_1}+1$ \\
		4.~~~~Obtain $\left( {{{{\bf{\tilde R}}}^{({r_1})}},{{\left\{ {{\bf{\tilde w}}_k^{({r_1})}} \right\}}_{k \in \mathbb{K}}}} \right)$ via the SDR/ZF algorithm.\\
		5.~~~~Update $\left( {{{{\bf{\tilde R}}}},{{\left\{ {{\bf{\tilde w}}_k} \right\}}_{k \in \mathbb{K}}}} \right) \leftarrow \left( {{{{\bf{\tilde R}}}^{({r_1})}},{{\left\{ {{\bf{\tilde w}}_k^{({r_1})}} \right\}}_{k \in \mathbb{K}}}} \right)$. \\
		6.~~~~Set ${r_2} \leftarrow 0$. \\
		7.~~~~{\bf{repeat}} \\
		8.~~~~~~~${r_2} \leftarrow {r_2}+1$ \\
		9.~~~~~~~Transform the non-convex problem P5 into P7. \\
		10.~~~~~~Compute the optimal solution of P7 via CVX. \\
		11.~~~~~~Update $\varepsilon _{\rm{e}}^{({r_2})}$ by solving P7 with $\varepsilon _{\rm{e}}^{({r_2}+1)}$. \\
		12.~~~{\bf{until}} $\left| {\varepsilon _{\rm{e}}^{({r_2} + 1)} - \varepsilon _{\rm{e}}^{({r_2})}} \right| \le {\iota _2}$. \\
		13.~~~Update $\left( {\left\{ {{\varepsilon _{{\rm{u}},k}}} \right\},{\varepsilon _{\rm{a}}},{\varepsilon _{\rm{p}}}} \right) \leftarrow \left( {\left\{ {\varepsilon _{{\rm{u}},k}^{({r_2} + 1)}} \right\},\varepsilon _{\rm{a}}^{({r_2} + 1)},\varepsilon _{\rm{p}}^{({r_2} + 1)}} \right)$. \\
		14.~~~Compute $C_{\rm{s}}^{({r_1})} = \mathop {\min }\limits_k {\log _2}\left( {1 + {\varepsilon _{{\rm{u}},k}}} \right) - \log 2\left( {1 + {\varepsilon _{\rm{a}}} + {\varepsilon _{\rm{p}}}} \right)$. \\
		15.~{\bf{until}} $\left| {C_{\rm{s}}^{({r_1} + 1)} - C_{\rm{s}}^{({r_1})}} \right| \le {\iota _1}$. \\
		{\bf{Output:}} $\left( {{\bf{\tilde R}},{{\left\{ {{{{\bf{\tilde w}}}_k}} \right\}}_{k \in \mathbb{K}}},\left\{ {{\varepsilon _{{\rm{u}},k}}} \right\},{\varepsilon _{\rm{a}}},{\varepsilon _{\rm{p}}}} \right)$. \\
		
		\bottomrule[1pt] 
	\end{tabular}
\end{table}
 \subsection{Convergence Analysis of the AO-TSS}
In the following, we analyze the convergence of {\bf Algorithm 2} including a two-stage optimization. For the first stage, i.e., step 4 to step 5 in {\bf Algorithm 2}, under the fixed SINR thresholds $\left( {\left\{ {{\varepsilon _{{\rm{u}},k}}} \right\},{\varepsilon _{\rm{a}}},{\varepsilon _{\rm{p}}}} \right)$, we can obtain the optimal beamformers ${{{\bf{\tilde R}}}}$ and ${{\left\{ {{\bf{\tilde w}}_k} \right\}}_{k \in \mathbb{K}}}$ by using the SDR/ZF algorithm. In the ${r_1}$-th iteration, although the SINR thresholds $ \left\{ {\varepsilon _{{\rm{u}},k}^{({r_1} - 1)}} \right\}$, $\varepsilon _{\rm{a}}^{({r_1} - 1)}$, $\varepsilon _{\rm{p}}^{({r_1} - 1)} $ influence the size of the feasible region of the first-stage optimization problem, ${{{\bf{\tilde R}}}^{({r_1}-1)}}$ and ${{\left\{ {{\bf{\tilde w}}_k^{({r_1}-1)}} \right\}}_{k \in \mathbb{K}}}$ are still in the feasible region. In other words, the optimal objective function value $L\left( {{{\bf{R}}^{({r_1})}},\delta _1^{({r_1})}} \right)$ of the first stage in the ${r_1}$-th iteration will not exceed that obtained in the $({r_1} - 1)$-th iteration.

Therefore, the sensing objective function is non-increasing over iterations as
\begin{align}
\label{con1}
L\left( {{{\bf{R}}^{({r_1})}},\delta _1^{({r_1})}} \right) \le L\left( {{{\bf{R}}^{({r_1}-1)}},\delta _1^{({r_1}-1)}} \right).
\end{align}
For the optimization in the second stage including the inner and outer iteration loop, i.e., step 6 to step 12 in {\bf Algorithm 2}, the SINR thresholds are designed to maximize the secrecy rate under the fixed optimal beamformers. In the inner loop, the optimized secrecy rate satisfies
\begin{align}
\label{con2}
C_{\rm{s}}^{({r_2})} \ge C_{\rm{s}}^{({r_2}-1)}.
\end{align}
For the outer loop, the optimal point sequence $\left( {\left\{ {\varepsilon _{{\rm{u}},k}^{({r_1})}} \right\},\varepsilon _{\rm{a}}^{({r_1})},\varepsilon _{\rm{p}}^{({r_1})}} \right)$ always results in a larger secrecy rate than the previous point sequence $\left( {\left\{ {\varepsilon _{{\rm{u}},k}^{({r_1} - 1)}} \right\},\varepsilon _{\rm{a}}^{({r_1} - 1)},\varepsilon _{\rm{p}}^{({r_1} - 1)}} \right)$. Thus, we have
\begin{align}
\label{con3}
C_{\rm{s}}^{({r_1})} \ge C_{\rm{s}}^{({r_1}-1)}.
\end{align}
As a result, the objective $L\left( {{{\bf{R}}},\delta _1} \right)$ decreases and $C_{\rm{s}}$ increases over iterations. Therefore, we can conclude that {\bf Algorithm 2} converges.

\section{Numerical Results} \label{5jie}                         
In this section, numerical results are provided to evaluate the proposed two-stage optimization problem for the secure ISAC system. The BS is equipped with a uniform linear array (ULA), where the number of antennas is ${\it{M}}=10$, and the antenna spacing is half wavelength, with the total transmit power 1 W. Angles at the radar are uniformly sampled with the resolution of 1$^\circ$ in the range of [$-90^\circ$, 90$^\circ$]. There are ${\it{Q}}=4$ targets located at angles $-60^\circ$, $-20^\circ$, $20^\circ$ and $60^\circ$, among which the target located at $60^\circ$ is identified as AE. We set the noise power for all LUs and Eves as $\sigma _{{\rm{c}}}^2 = \sigma _{\rm{s}}^2 = \sigma _{\rm{p}}^2 = 0.01~{\rm{W}}$. The interference power from the AE to LUs is fixed as 0.01 W. The received SINR threshold of each LU equals ${\varepsilon _{{\rm{u}}}}={\varepsilon _{{\rm{u}},k}},\forall k$. For the PE, the probability parameter $\tau $ is set as 0.95. Each element of the $k$-th channel vector ${{\bf{h}}_{\it{k}}}$ is assumed to obey complex Gaussian distribution, i.e. ${{\bf{h}}_{\it{k}}}\sim \mathcal{CN}(0,{{\bf{I}}_M})$. The channel from the AE to the $k$-th LU is modeled as ${{\it{h}}_{{\rm{a}},{\it{k}}}}\sim \mathcal{CN}(0,1)$. Additionally, we consider that the PE is located farther away from the BS than LUs, and the variance of each entry in ${{\bf{h}}_{\rm{p}}}$ is 0.001.\footnote{Due to the leakage during the baseband conversion of the received signal, the PE may be detected and removed by the BS. Thus, the PE may choose a location far from the BS to reduce the probability of being detected, resulting in greater signal attenuation compared with LUs.}

 Besides, we compare the proposed algorithms with a baseline scheme in \cite{19}. The only secure threat in \cite{19} comes from targets disguised by PEs. As a result, the baseline scheme is insufficient to resist multiple hybrid-colluding Eves, leading to a poor secrecy rate. However, under the same resource settings, it may achieve higher sensing performance.
	
	In the subsequent simulations, we first evaluate the radar performance under perfect AE CSI and statistical PE CSI in Subsection V-A and further analyze the radar performance under relaxed CSI in Subsection V-B. Then, the communication performance is validated in Subsection V-C.
\subsection{Radar Performance Evaluation under Perfect AE CSI and Statistical PE CSI}                     
\begin{figure}[htbp]
\centering
\includegraphics[width=3.2in]{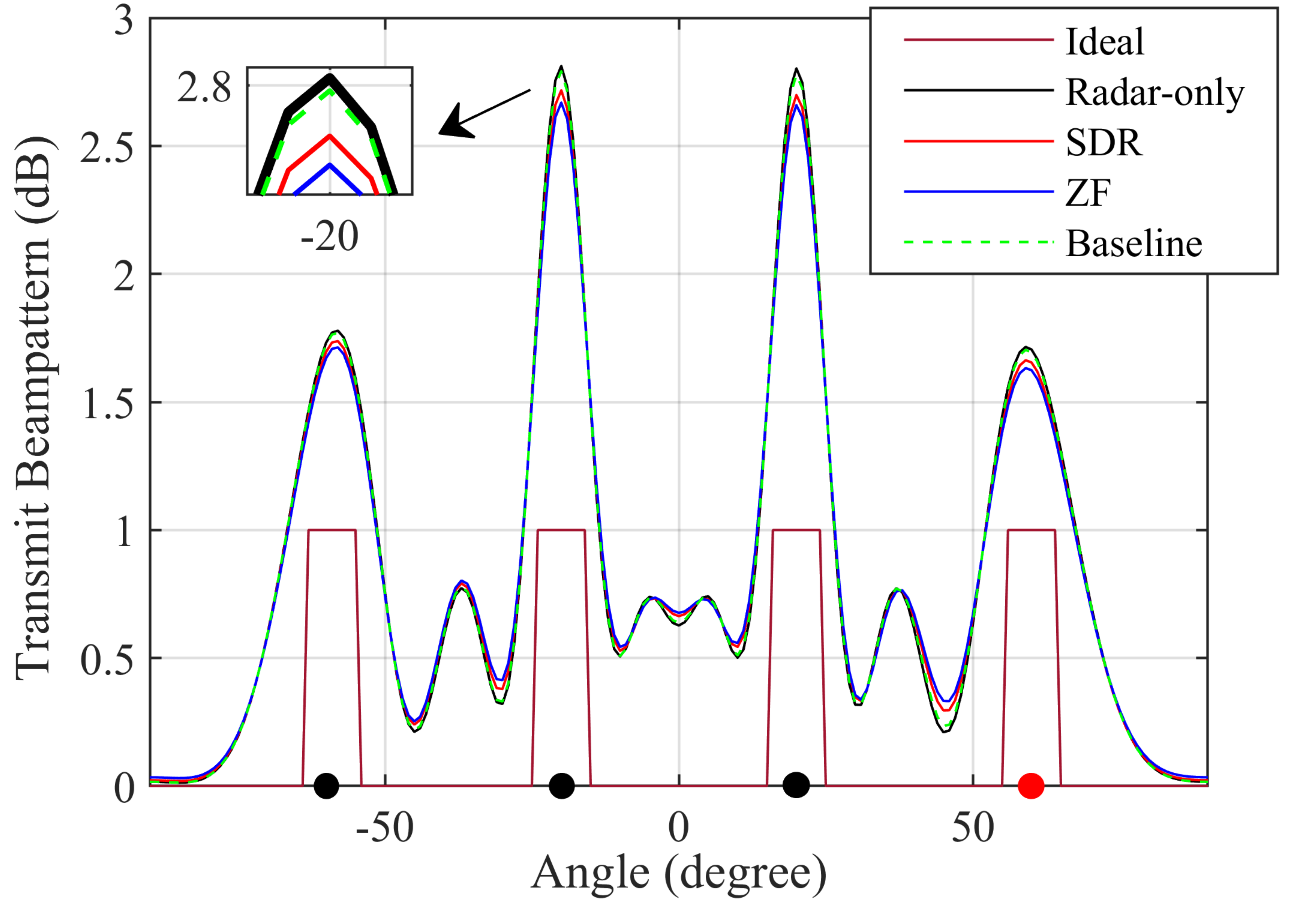}
\caption{The transmit beampatterns with angles with respect to ${\it{K}} = 2$, ${\varepsilon _{{\rm{u}}}} = 16~{\rm{dB}}$, ${{\varepsilon _{\rm{a}}}} = 2~{\rm{dB}}$ and ${{\varepsilon _{\rm{p}}}} = 2~{\rm{dB}}$.}
\label{beampatterm}
\end{figure}

Fig. \ref{beampatterm} displays the two-dimensional\footnote{The array elements of ULA are uniformly arranged in a straight line, and the array can adjust the beam direction in a two-dimensional plane.} beampatterns of the proposed SDR and ZF algorithms, in which the SINR thresholds are set as ${\varepsilon _{{\rm{u}}}} = 16~{\rm{dB}}$, ${{\varepsilon _{\rm{a}}}} = 2~{\rm{dB}}$, ${{\varepsilon _{\rm{p}}}} = 2~{\rm{dB}}$. The solid circles indicate the target angles, among which the red one denotes the AE. To reveal more insights, we also plot waveform diagrams obtained by the baseline. It is observed that the proposed SDR and ZF algorithms are capable of forming beampatterns which are close to the radar-only beampattern. In particular, the mainlobe power of the ZF beamforming is lower than that of SDR beamforming, implying the sensing performance loss of ZF beamforming compared with SDR beamforming. Although the proposed algorithms impose a slightly performance degradation on the transmit beampattern compared to the baseline, the secure communication can still be guaranteed, especially in the complex hybrid-colluding eavesdropping scenarios.

\begin{figure}[htbp]
\centering
\includegraphics[width=3.2in]{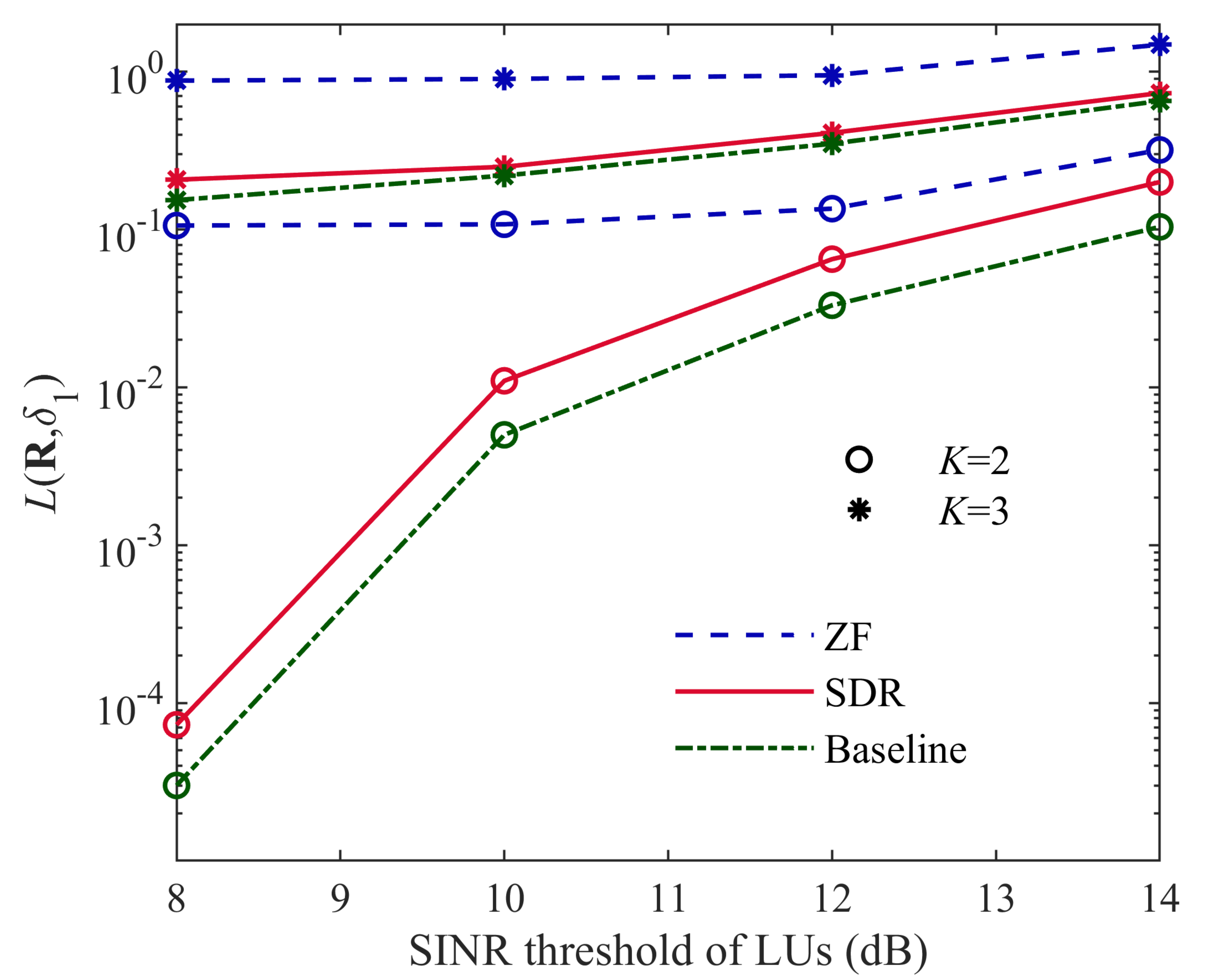}
\caption{The sensing objective $L({\bf{R}},{\delta _1})$ with respect to the SINR threshold of LUs under different number of LUs $K$ ($M$ = 10, ${{\varepsilon _{\rm{a}}}} = 2~{\rm{dB}}$ and ${{\varepsilon _{\rm{p}}}} = 2~{\rm{dB}}$).}
\label{sensing_objective_L_b}
\end{figure}

Fig. \ref{sensing_objective_L_b} plots the sensing performance objective $L({\bf{R}},{\delta _1})$ with respect to the SINR threshold of LUs. All simulation results represent averaged values over 1000 Monte Carlo trials. In Fig. \ref{sensing_objective_L_b}, we can observe that $L({\bf{R}},{\delta _1})$ of each algorithm increases with ${{\varepsilon _{\rm{u}}}}$, implying that the increased communication requirements deteriorate the sensing performance. Specifically, the sensing curve of ZF algorithm experiences a slight increase. This is because the ZF algorithm can eliminate MUI and radar interference, resulting in a higher SINR. Moreover, the SDR algorithm achieves improved sensing performance compared with the ZF algorithm, and the sensing performances of the SDR and ZF algorithms tend to become close for a high ${\varepsilon _{{\rm{u}}}}$. In addition, the baseline consistently exhibits a slightly low $L({\bf{R}},{\delta _1})$ because it only considers the scenario where the targets are potential PEs, so that the security of information is not fully taken into account. Besides, Fig. \ref{sensing_objective_L_b} also demonstrates the impact of the number of LUs $K$ on sensing performance. It can be found that the more users the system serves, the higher $L({\bf{R}},{\delta _1})$ becomes. Compared with the SINR threshold of LUs, the number of LUs $K$ has a greater impact on the sensing performance. This implies that serving more downlink users is more restrictive than increasing the LU SINR level.

\begin{figure}[htbp]
\centering
\includegraphics[width=3.2in]{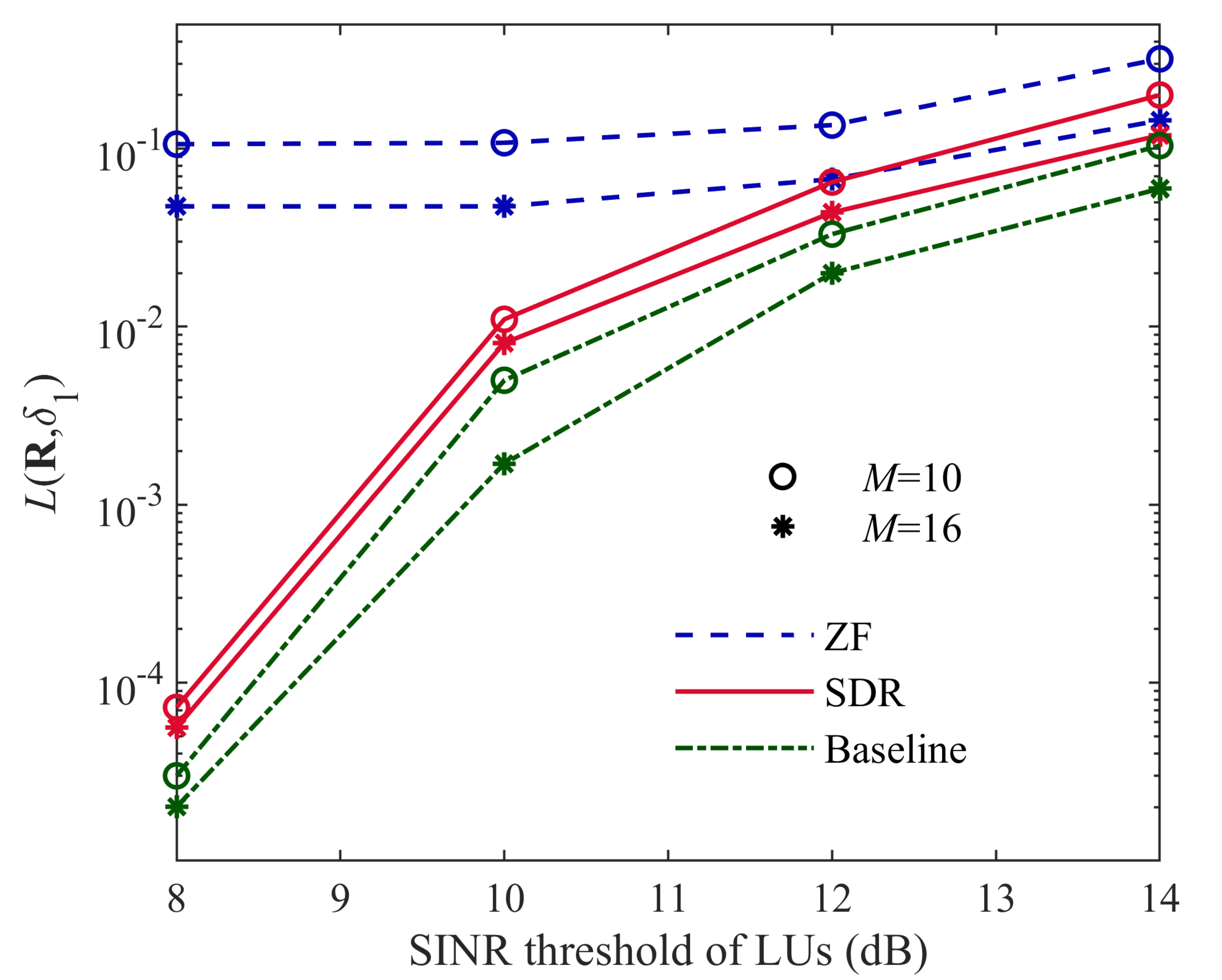}
\caption{The sensing objective $L({\bf{R}},{\delta _1})$ with respect to the SINR threshold of LUs under different number of transmit antennas $M$ ($K$ = 2, ${{\varepsilon _{\rm{a}}}} = 2~{\rm{dB}}$ and ${{\varepsilon _{\rm{p}}}} = 2~{\rm{dB}}$).}
\label{L_M}
\end{figure}

 Fig. \ref{L_M} plots the sensing performance objective $L({\bf{R}},{\delta _1})$ with respect to the SINR threshold of LUs for $M$ = 10 and 16. Similar to Fig. \ref{sensing_objective_L_b}, as the SINR threshold increases, the sensing performance deteriorates. Additionally, the sensing performance with 16 transmit antennas is better than with 10 antennas.

\begin{figure}[htbp]
\centering
\includegraphics[width=2.8in]{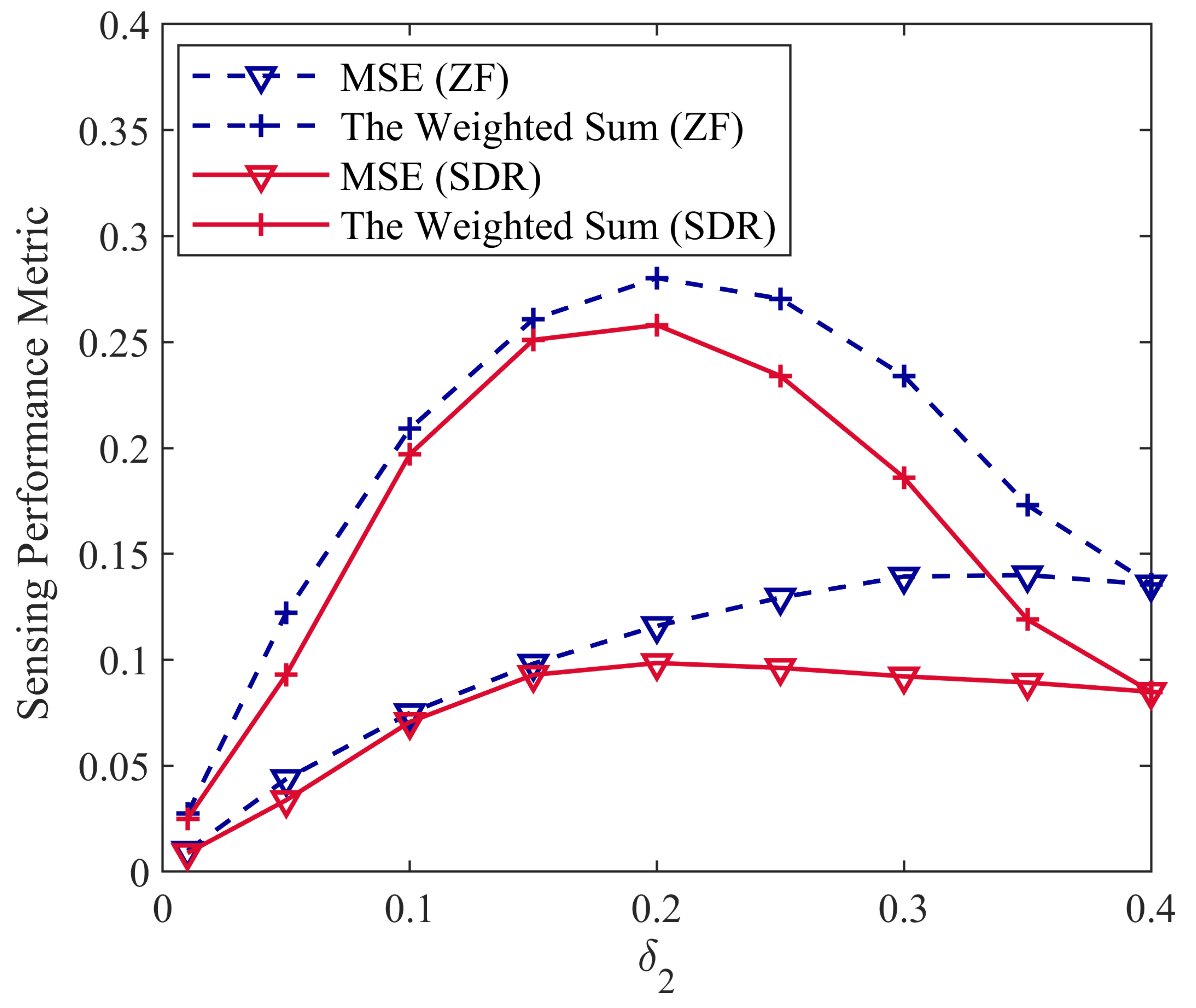}
\caption{The sensing performance metric with respect to the weighting factor ${\delta _2}$  $({\varepsilon _{{\rm{u}}}} = 12~{\rm{dB}}$, ${{\varepsilon _{\rm{a}}}} = 2~ {\rm{dB}}$, ${{\varepsilon _{\rm{p}}}} = 2 ~{\rm{dB}})$.}
\label{weighted_sum}
\end{figure}

Fig. \ref{weighted_sum} presents the sensing performance of the SDR and ZF algorithms including the beampattern MSE, the weighted sum $L({\bf{R}},{\delta _1})$ with respect to the weighting factor ${\delta _2}$. It can be found that with the increase of ${\delta _2}$, the beampattern MSE increases, while the weighted sum increases first and then decreases. When ${\delta _2}=0$ or ${\delta _2}=0.4$, the weighted sum coincides with beampattern MSE. Besides, no matter what ${\delta _2}$ is, the weighted sum values of the ZF algorithm are higher than the SDR algorithm. This shows that under the same simulation conditions, the SDR algorithm can achieve better sensing performance than the ZF algorithm, although the latter has the advantage of reducing computational complexity.

 Furthermore, we also show the actual angle estimation performance of multiple targets at the BS to validate the efficiency of our proposed joint transmit beamforming design. In the simulation, the angles $\left\{ {{{\theta }_q}} \right\}_{q = 1}^Q$ are estimated based on the received signals ${\bf{y}}\left( n \right) = {\bf{Ax}}\left( n \right) + {{\bf{n}}_{\rm{y}}}$ via the maximum likelihood estimation (MLE) technique \cite{MLE}, in which the target response matrix equals ${\bf{A}} = \sum\nolimits_{q = 1}^Q {{\bf{a}}\left( {{\theta _q}} \right){{\bf{a}}^{\rm{H}}}\left( {{\theta _q}} \right)} $. Specifically, the estimation of the angles is obtained by maximizing the log-likelihood function $ {\rm{tr}}\left( {{{\bf{P}}_{{\bf{A}}}}{{\bf{R}}_y}} \right)$, where ${{\bf{P}}_{{\bf{A}}}} = {\bf{A}}{\left( {{{\bf{A}}^{\rm{H}}}{\bf{A}}} \right)^{ - 1}}{{\bf{A}}^{\rm{H}}}$ is the projection operator onto the space spanned by the columns of the matrix ${\bf{A}}$, and ${{\bf{R}}_y} = \frac{1}{N}\sum\nolimits_{n = 1}^N {{\bf{y}}\left( n \right){{\bf{y}}^{\rm{H}}}\left( n \right)} $ represents the sample covariance matrix. Moreover, the angle estimation performance is evaluated by the root mean squared error (RMSE), defined as
\begin{align}
\label{RMSE}
RMSE = \sqrt {\frac{1}{Q}\sum\nolimits_{q = 1}^Q {{{\left( {{{\tilde \theta }_q} - {\theta _q}} \right)}^2}} } ,
\end{align}
where ${{{\tilde \theta }_q}} $ denotes the estimate of $ {{\theta _q}} $.

\begin{figure}[htbp]
\centering
\includegraphics[width=2.8in]{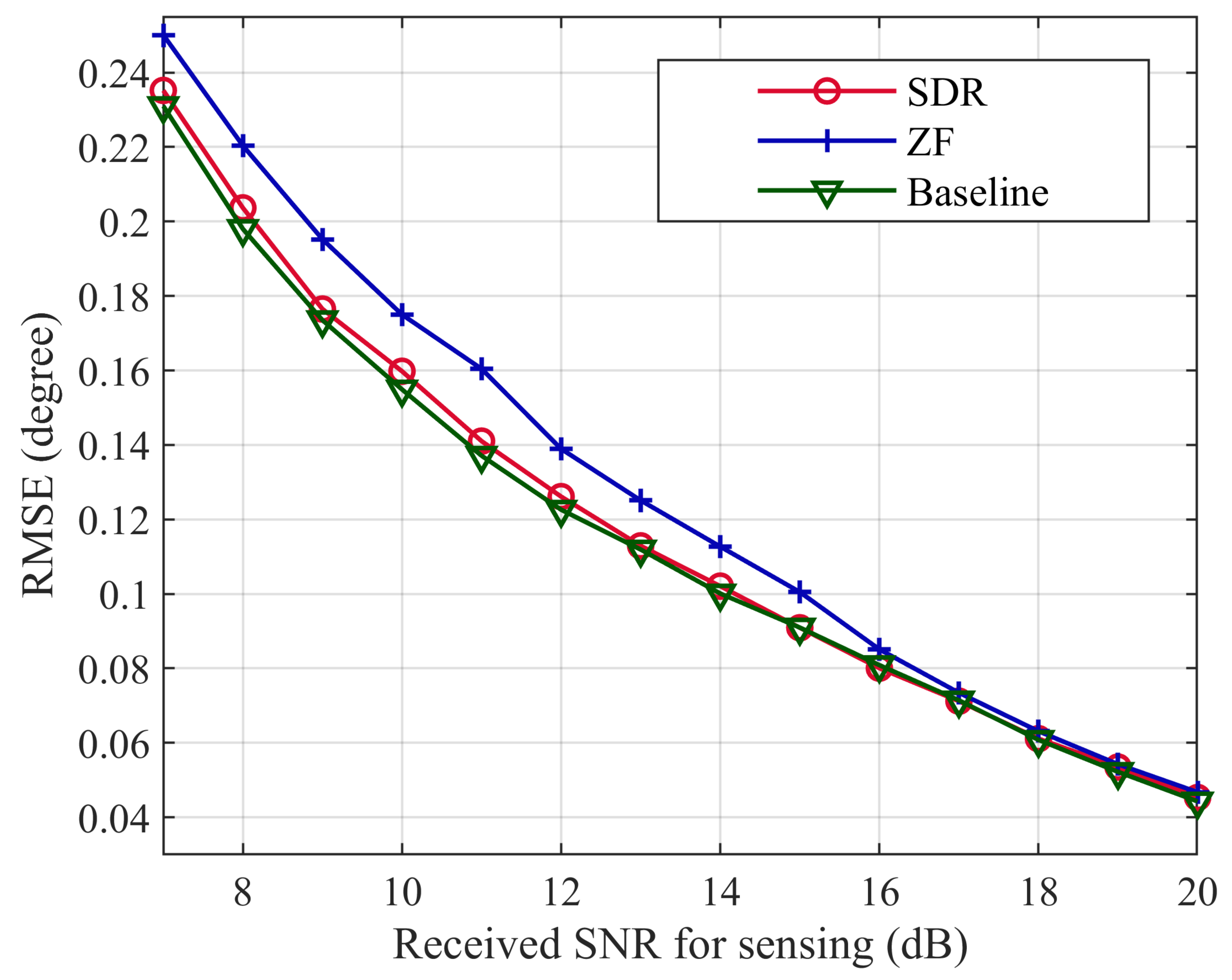}
\caption{RMSE for angle estimation with respect to the received SNR for sensing.}
\label{DOA_MLE}
\end{figure}

In Fig. \ref{DOA_MLE}, the RMSE curves are illustrated for different SNR at the BS, with $N$ = 100 and 500 Monte Carlo trials. It is shown that the RMSE decreases with increasing SNR and the ZF algorithm performs worse than the SDR algorithm and the baseline, especially in the low SNR region. Furthermore, we can observe that the baseline obtains slightly better multi-target angle estimation performance, since the PLS aspects of confidential information protection is not fully taken into account in this algorithm.

 \subsection{Radar Performance Evaluation under Relaxed CSI Knowledge}
\begin{figure}[htbp]
\centering
\includegraphics[width=2.8in]{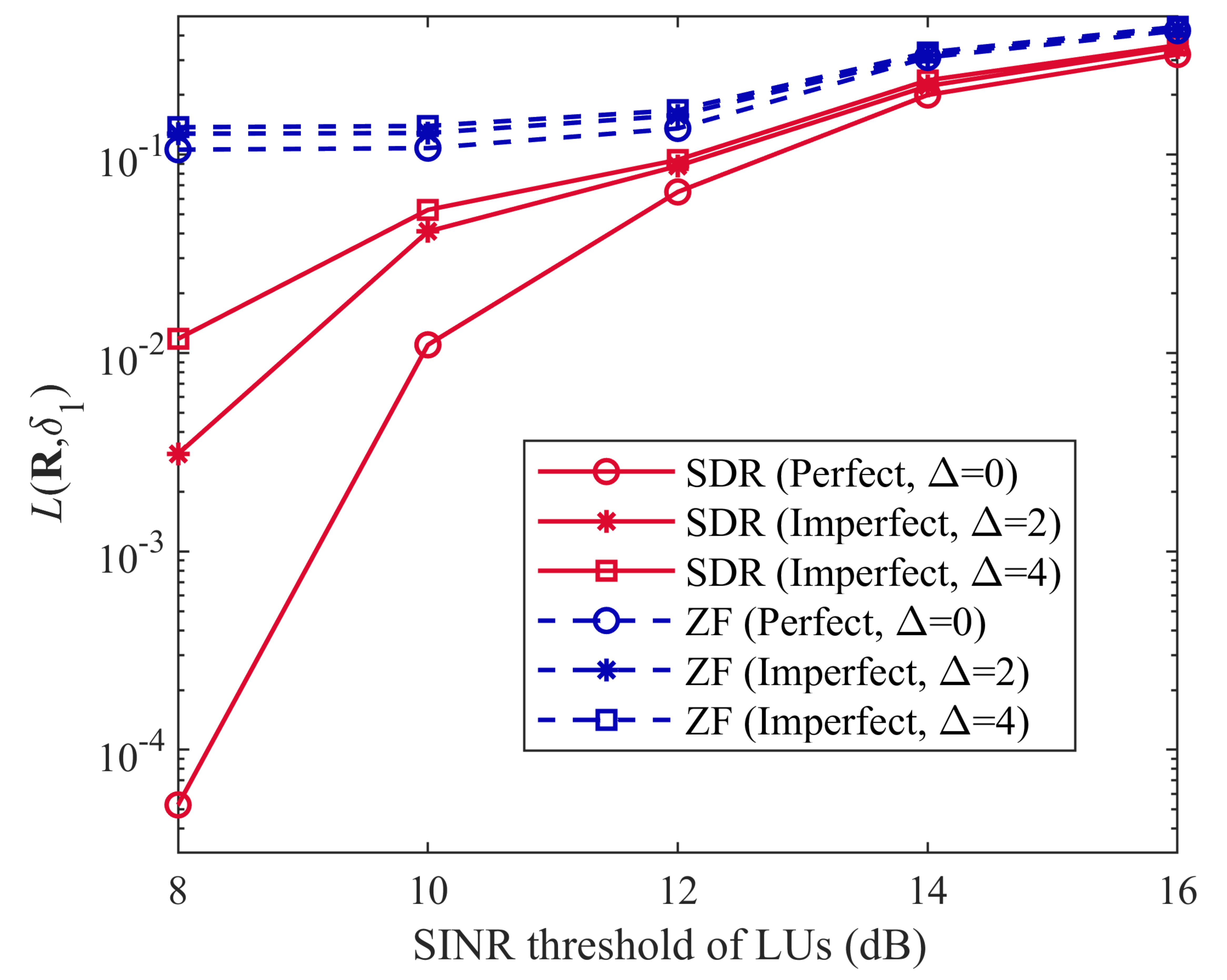}
\caption{The sensing objective $L({\bf{R}},{\delta _1})$ comparison with different angular uncertainties of the AE ($K=2$, ${{\varepsilon _{\rm{u}}}} = 12 ~{\rm{dB}}$, ${{\varepsilon _{\rm{a}}}} = 2~ {\rm{dB}}$, ${{\varepsilon _{\rm{p}}}} = 2 ~{\rm{dB}})$.}
\label{AE_robutness}
\end{figure}

In Fig. \ref{AE_robutness}, we evaluate the impact of the direction uncertainties on the optimization performance. The SINR threshold of LUs and the Eves are set to ${{\varepsilon _{\rm{u}}}} = 12 ~{\rm{dB}}$, ${{\varepsilon _{\rm{a}}}} = 2~ {\rm{dB}}$, ${{\varepsilon _{\rm{p}}}} = 2 ~{\rm{dB}}$, respectively. \enquote{Perfect, $\Delta=0$} means the case of perfect CSI of AE. We can observe that the increase in uncertainty of the AE introduces additional constraints to the optimization problem, consequently diminishing sensing performance. As demonstrated in Fig. \ref{AE_robutness}, the sensing objective increases with the AE uncertainty interval. In addition, the performance gaps between the two algorithms notably become smaller at high LU SINR constraints.

\begin{figure}[htbp]
\centering
\includegraphics[width=2.8in]{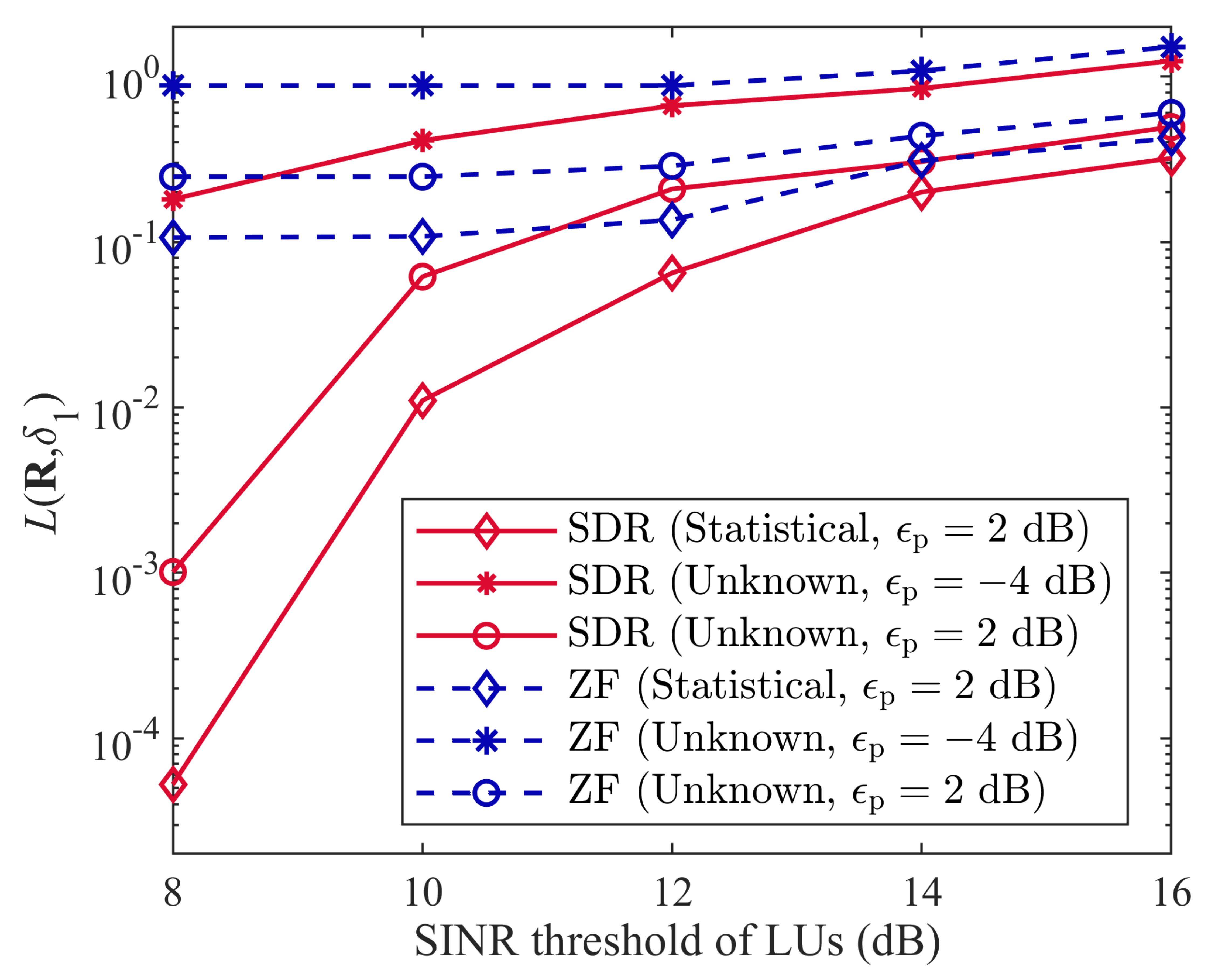}
\caption{The sensing objective $L({\bf{R}},{\delta _1})$ comparison with unknown CSI of the PE ($K=2$, ${{\varepsilon _{\rm{u}}}} = 12 ~{\rm{dB}}$, ${{\varepsilon _{\rm{a}}}} = 2~ {\rm{dB}}$).}
\label{PE_robutness}
\end{figure}

Fig. \ref{PE_robutness} presents the performance of the proposed joint transmit beamforming design under different settings of ${{\varepsilon _{\rm{p}}}}$ when CSI of PE is unknown. \enquote{Statistical, $\epsilon_{\rm{p}}=2$ dB} means the case of the statistical CSI of PE with $\epsilon_{\rm{p}}=2$ dB. Not surprisingly, it can be observed that the sensing performance degrades significantly due to the anti-eavesdropping design for multiple angles. Furthermore, as the SINR ${{\varepsilon _{\rm{p}}}}$ of PE decreases, the secure performance enhances at the expense of deteriorating sensing performance.

\subsection{Secure Communication Performance}                    

\begin{figure}[htbp]
\centering
\includegraphics[width=2.8in]{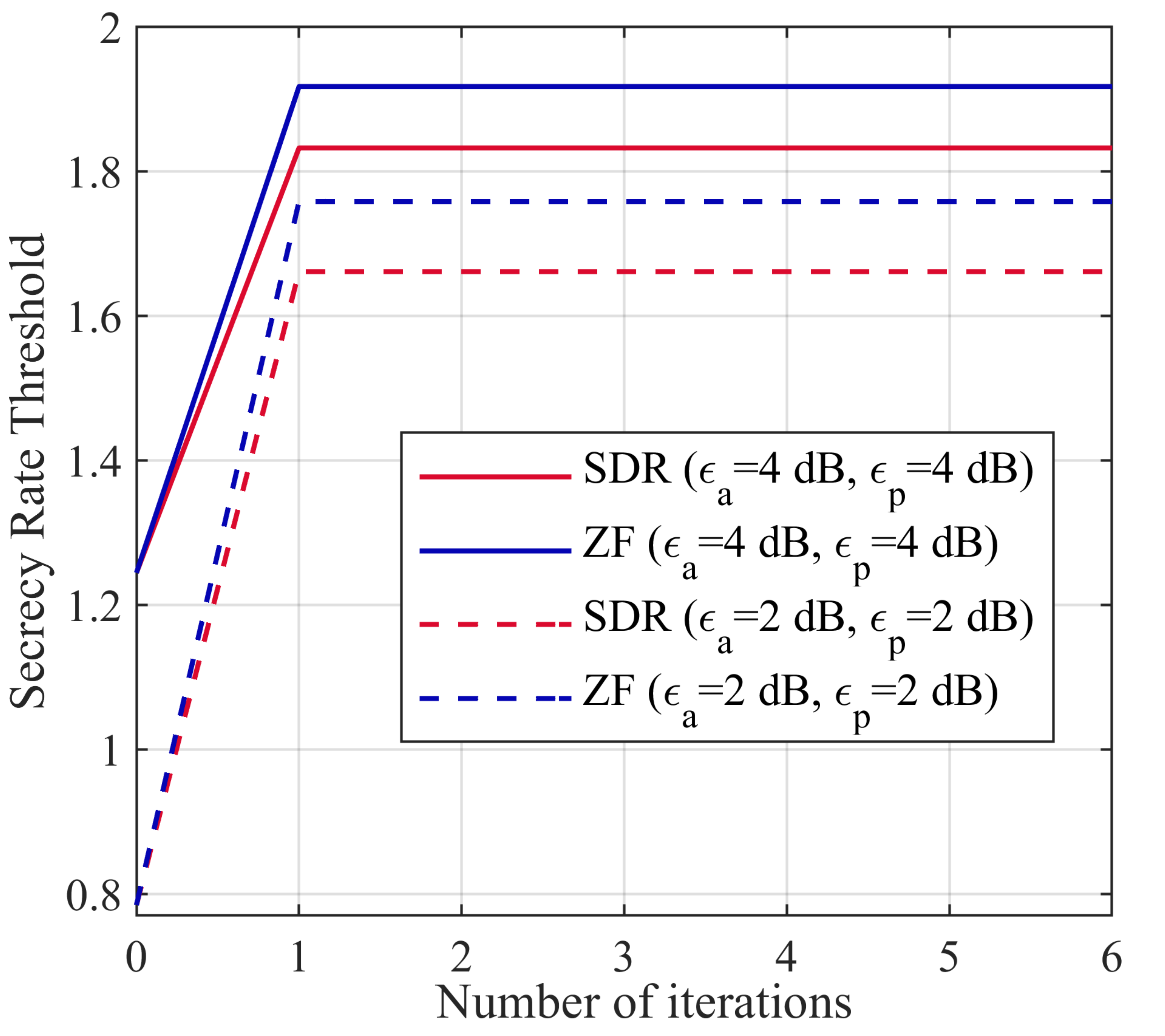}
\caption{The secrecy rate thresholds with respect to the number of iterations with different SINR thresholds of Eves.}
\label{convergence}
\end{figure}

We validate the effectiveness of the algorithm for maximizing the secrecy rate threshold, which is the objective function of the second-stage problem. The convergence rate is given in Fig. \ref{convergence}. Obviously, for different system parameters, the secrecy rate thresholds increase quickly and remain constant, indicating fast convergence. Furthermore, when the first-stage problem is solved with the ZF algorithm, the converged secrecy rate in the second-stage problem is higher than the SDR algorithm. This is because the interference is eliminated and the feasible region of the SINR thresholds is extended.

\begin{figure}[htbp]
\centering
\includegraphics[width=2.8in]{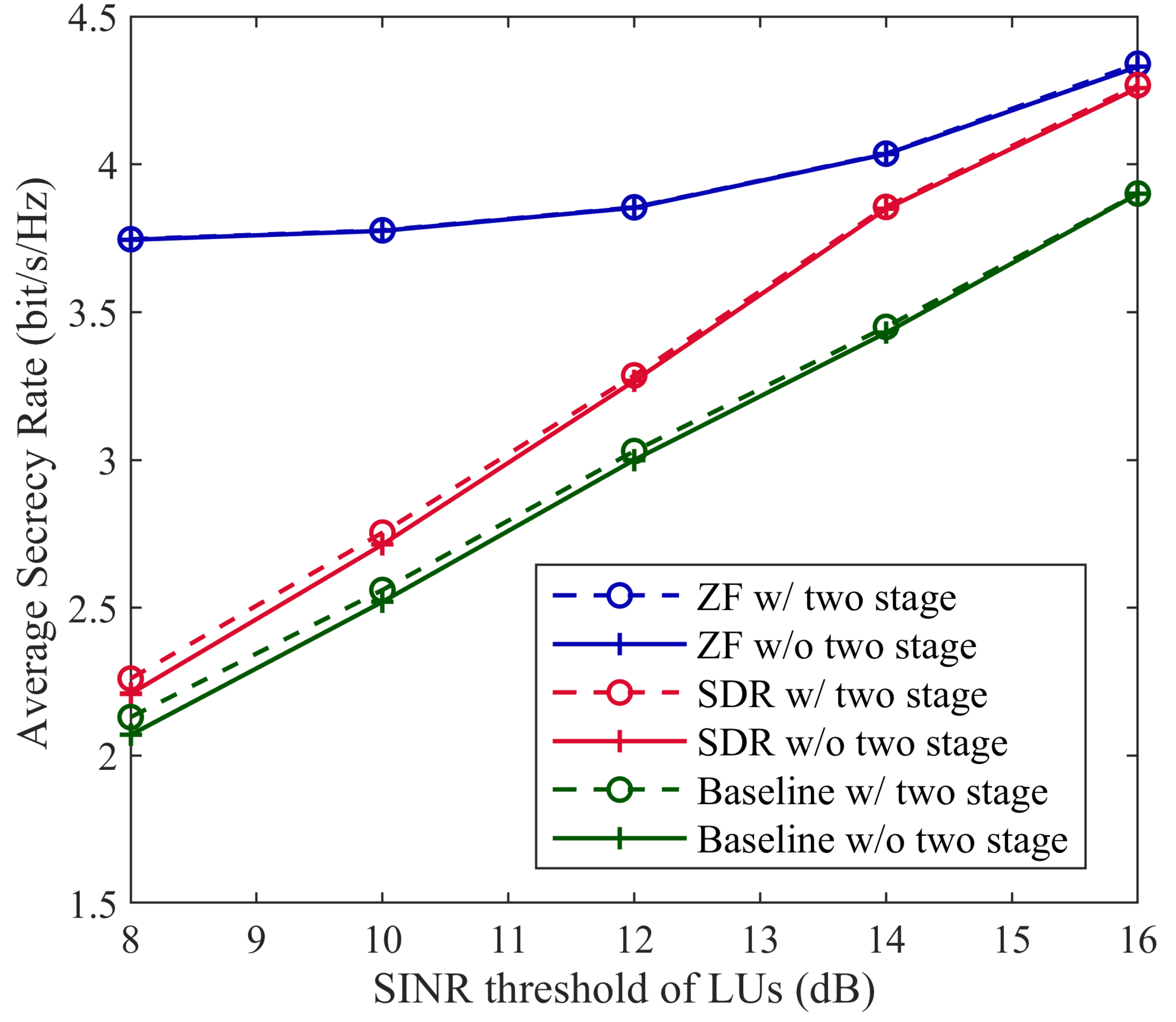}
\caption{The average system secrecy rate with respect to the SINR threshold of LUs  $({{\varepsilon _{\rm{a}}}} = 2~{\rm{dB}}$, ${{\varepsilon _{\rm{p}}}} = 2~{\rm{dB}}$, $K=2$).}
\label{fig_6}
\end{figure}

We further plot the average system secrecy rate with respect to the SINR threshold of LUs in Fig. \ref{fig_6}. As we can see, the average system secrecy rate increases as the SINR threshold grows, and the proposed two-stage scheme achieves a higher secrecy rate compared with the scheme without two stage, especially at low SINR threshold. However, this enhancement becomes less conspicuous as the threshold increases, primarily due to the reduction in the feasible region of optimization variables. Furthermore, due to the elimination of interference, the ZF algorithm achieves higher average secrecy rates than the SDR algorithm, while the performance of the SDR and ZF algorithms tends to become similar at high SINR. Besides, the secure performance of the baseline is relatively poor, because it only focuses on the simple eavesdropping scenarios where the PEs disguise themselves as targets, which is insufficient for the case of multiple hybrid-colluding Eves.

Table \ref{diff1} shows the sensing gaps between the scheme with and without two stage. It can be seen that under each SINR threshold, the gaps are negligible. From Fig. \ref{fig_6} and Table \ref{diff1}, we can conclude that the secure communication performance of the proposed two-stage problem can be improved under the condition of ensuring the optimal radar performance.
\begin{table}[H]
\belowrulesep=0pt
\aboverulesep=0pt
\renewcommand\arraystretch{1.5}
\caption{\textbf{The sensing gaps between the scheme with and without two stage}}
\label{diff1}
\centering
\begin{tabular}{c|cccc}
\toprule
${\varepsilon _{\rm{u}}}$&6$~{\rm{~(dB)}}$&10$~{\rm{~(dB)}}$&14$~{\rm{~(dB)}}$ \\
\midrule
SDR&$3.27\times {10^{ - 5}}$&$7.56\times {10^{ - 5}}$&$9.44\times {10^{ - 5}}$ \\
ZF&$7.99\times {10^{ - 5}}$&$7.65\times {10^{ - 5}}$&$5.02\times {10^{ - 5}}$\\
Baseline&$4.51\times {10^{ - 5}}$&$3.02\times {10^{ - 5}}$&$7.49\times {10^{ - 5}}$ \\
\bottomrule
\end{tabular}
\end{table}

\section{Conclusion}\label{6jie}                  
This paper has studied the PLS problem for ISAC system that consists of a BS, multiple LUs, an external PE and multiple sensing targets, where one sensing target is disguised as the AE. We have considered the case that AE and PE collude, and proposed the AO-TSS to optimize the system performance. In the first stage, we have assumed that the BS could obtain the perfect AE CSI and statistical PE CSI. Based on this, we have formulated the non-convex optimization problem to ensure the accuracy of target sensing and security of information transmission by jointly designing the information and sensing beamforming, while meeting the secure communication constraints. Then, we have proposed the SDR algorithm and reduced complexity ZF algorithm to solve the non-convex problem. Besides, we have also designed the beamformers for the cases of imperfect AE CSI and unknown PE CSI, respectively. Moreover, we have formulated the second-stage max-min optimization problem to enhance the system secure performance while ensuring the radar performance. Finally, the superior performance in the sensing and secure communication has been illustrated via extensive numerical results.
\appendices
\section*{Appendix A}                            
By exploiting some mathematical manipulations, the probability in (17d) can be rewritten as
\begin{align}
\label{46}
\begin{array}{l}
{\rm{Pr}}\left( {\dfrac{{{{\bf{h}}_{\rm{p}}^{\rm{H}}}{{\bf{R}}_{\rm{c}}}{{\bf{h}}_{\rm{p}}}}}{{{{\bf{h}}_{\rm{p}}^{\rm{H}}}{{\bf{R}}_{\rm{s}}}{{\bf{h}}_{\rm{p}}} + {\sigma ^2_{\rm{p}}}}} \le {\varepsilon _{\rm{p}}}} \right)
 = {\rm{Pr}}\left( {\dfrac{{{\rm{Tr(}}{{\bf{H}}_{\rm{p}}}{{\bf{R}}_{\rm{c}}})}}{{{\rm{Tr(}}{{\bf{H}}_{\rm{p}}}{{\bf{R}}_{\rm{s}}}) + {\sigma ^2_{\rm{p}}}}} \le {\varepsilon _{\rm{p}}}} \right),
\end{array}
\end{align}
where ${{\bf{H}}_{\rm{p}}} = {\bf{h}}_{\rm{p}}^{\rm{H}}{{\bf{h}}_{\rm{p}}}$ with ${{\bf{h}}_{\rm{p}}} \sim \mathcal{CN}({\bf{0}},c{{\bf{I}}_M})$. Thus, constraint (17d) is expressed as
\begin{align}
\label{47}
\begin{array}{l}
\Pr \left\{ {\dfrac{{{\rm{Tr(}}{{\bf{H}}_{\rm{p}}}{{\bf{R}}_{\rm{c}}})}}{{{\rm{Tr(}}{{\bf{H}}_{\rm{p}}}{{\bf{R}}_{\rm{s}}}) + {\sigma ^2_{\rm{p}}}}} \le {\varepsilon _{\rm{p}}}} \right\} \ge \tau.
\end{array}
\end{align}
Then, (\ref{47}) can be further expressed as
\begin{align}
\label{471}
\begin{array}{l}
\Pr \left\{ {{\rm{Tr}}\left( {{{\bf{H}}_{\rm{p}}}\left( {{{\bf{R}}_{\rm{c}}} - {\varepsilon _{\rm{p}}}{{\bf{R}}_{\rm{s}}}} \right)} \right) \le {\varepsilon _{\rm{p}}}{\sigma ^2_{\rm{p}}}} \right\} \ge \tau.
\end{array}
\end{align}
The probability in (\ref{471}) cannot be computed directly unless specific properties of ${{{\bf{H}}_{\rm{p}}}\left( {{{\bf{R}}_{\rm{c}}} - {\varepsilon _{\rm{p}}}{{\bf{R}}_{\rm{s}}}} \right)}$ are satisfied. We then replace the probability constraint to its upper bound as
\begin{align}
\label{48}
\begin{array}{l}
{\rm{Tr}}({{\bf{H}}_{\rm{p}}}({{\bf{R}}_{\rm{c}}} - {\varepsilon _{\rm{p}}}{{\bf{R}}_{\rm{s}}}))\mathop  \le \limits^{({\rm{a}})} \sum\limits_{m = 1}^M {{\lambda _m}({{\bf{H}}_{\rm{p}}}){\lambda _m}({{\bf{R}}_{\rm{c}}} - {\varepsilon _{\rm{p}}}{{\bf{R}}_{\rm{s}}})} \\
{\rm{                            }}\mathop  = \limits^{({\rm{b}})} {\lambda _{\max }}({{\bf{H}}_{\rm{p}}}){\lambda _{\max }}({{\bf{R}}_{\rm{c}}} - {\varepsilon _{\rm{p}}}{{\bf{R}}_{\rm{s}}})\\
{\rm{                            }}\mathop  = \limits^{({\rm{c}})} {\rm{Tr}}({{\bf{H}}_{\rm{p}}}){\lambda _{\max }}({{\bf{R}}_{\rm{c}}} - {\varepsilon _{\rm{p}}}{{\bf{R}}_{\rm{s}}}),
\end{array}
\end{align}
where ${\lambda _m}( \cdot )$ is the {\it{m}}-th eigenvalue of the matrix, and its orders are arranged as ${\lambda _{\max }}( \cdot ) = {\lambda _1}( \cdot ) \ge  \cdots  \ge {\lambda _M}( \cdot )$. In addition, (a) is from the properties of positive Hermitian matrix. (b) and (c) are  obtained because ${{\bf{H}}_{\rm{p}}}$ is a semi-definite Hermitian matrix of rank-1. Subsequently, the condition for ${{\bf{R}}_{\rm{c}}} - {\varepsilon _{\rm{p}}}{{\bf{R}}_{\rm{s}}}$ to be a semi-definite matrix can be omitted. Substituting (\ref{48}) into (\ref{471}), we obtain the inequality as
\begin{align}
\label{49}
\begin{array}{l}
\Pr \left\{ {{\rm{Tr}}\left( {{{\bf{H}}_{\rm{p}}}\left( {{{\bf{R}}_{\rm{c}}} - {\varepsilon _{\rm{p}}}{{\bf{R}}_{\rm{s}}}} \right)} \right) \le {\varepsilon _{\rm{p}}}{\sigma ^2_{\rm{p}}}} \right\}\\
 \hspace{10mm}\ge \Pr \left\{ {{\rm{Tr}}({{\bf{H}}_{\rm{p}}}){\lambda _{\max }}({{\bf{R}}_{\rm{c}}} - {\varepsilon _{\rm{p}}}{{\bf{R}}_{\rm{s}}}) \le {\varepsilon _{\rm{p}}}{\sigma ^2_{\rm{p}}}} \right\}.
\end{array}
\end{align}
Then, we have
\begin{align}
\label{5000}
\begin{array}{l}
\Pr \left\{ {\dfrac{{{\rm{Tr(}}{{\bf{H}}_{\rm{p}}}{{\bf{R}}_{\rm{c}}})}}{{{\rm{Tr(}}{{\bf{H}}_{\rm{p}}}{{\bf{R}}_{\rm{s}}}) + {\sigma ^2_{\rm{p}}}}} \le {\varepsilon _{\rm{p}}}} \right\}\\
{\rm{    }} \ge \Pr \left\{ {{\rm{Tr}}({{\bf{H}}_{\rm{p}}}){\lambda _{\max }}({{\bf{R}}_{\rm{c}}} - {\varepsilon _{\rm{p}}}{{\bf{R}}_{\rm{s}}}) \le {\varepsilon _{\rm{p}}}{\sigma ^2_{\rm{p}}}} \right\} \ge \tau.
\end{array}
\end{align}
Next, we introduce ${{{\bf{\mathord{\buildrel{\lower3pt\hbox{$\scriptscriptstyle\smile$}}
\over H} }}}_{\rm{p}}} = {{{\bf{\mathord{\buildrel{\lower3pt\hbox{$\scriptscriptstyle\smile$}}
\over h} }}}_{\rm{p}}}{\bf{\mathord{\buildrel{\lower3pt\hbox{$\scriptscriptstyle\smile$}}
\over h} }}_{\rm{p}}^{\rm{H}}$ with ${{{\bf{\mathord{\buildrel{\lower3pt\hbox{$\scriptscriptstyle\smile$}}
\over h} }}}_{\rm{p}}}\sim \mathcal{CN}({\bf{0}},{{\bf{I}}_M})$. Utilizing the the positive definiteness of the matrix ${{\bf{H}}_{\rm{p}}}$, we obtain the inequality as
\begin{align}
\label{50001}
\Pr \left\{ {\dfrac{1}{{{\rm{Tr}}({{\bf{H}}_{\rm{p}}})}} \le \dfrac{{{\lambda _{\max }}({{\bf{R}}_{\rm{c}}} - {\varepsilon _{\rm{p}}}{{\bf{R}}_{\rm{s}}})}}{{{\varepsilon _{\rm{p}}}\sigma _{\rm{p}}^2}}} \right\} \le 1 - \tau ,
\end{align}
where the trace of ${{\bf{H}}_{\rm{p}}}$ can be expanded as
\begin{align}
\label{50}
\begin{array}{l}
{\rm{Tr}}({{\bf{H}}_{\rm{p}}}) = h_{{\rm{p}},1}^2 + h_{{\rm{p}},2}^2 +  \cdots  + h_{{\rm{p}},M}^2\\
 = c [\dfrac{{h_{{\rm{p}},1}^2}}{c} + \dfrac{{h_{{\rm{p}},2}^2}}{c} +  \cdots  + \dfrac{{h_{{\rm{p}},M}^2}}{c}]\\
 = c (\mathord{\buildrel{\lower3pt\hbox{$\scriptscriptstyle\smile$}}
\over h} _{{\rm{p}},1}^2 + \mathord{\buildrel{\lower3pt\hbox{$\scriptscriptstyle\smile$}}
\over h} _{{\rm{p}},2}^2 +  \cdots  + \mathord{\buildrel{\lower3pt\hbox{$\scriptscriptstyle\smile$}}
\over h} _{{\rm{p}},M}^2)\\
 = c  {\rm{Tr}}({{{\bf{\mathord{\buildrel{\lower3pt\hbox{$\scriptscriptstyle\smile$}}
\over H} }}}_{\rm{p}}}).
\end{array}
\end{align}
Thus, (\ref{50001}) can be further converted to
\begin{align}
\label{500}
\begin{array}{l}
\Pr \left\{ {\dfrac{1}{{{\rm{Tr}}({{{\bf{\mathord{\buildrel{\lower3pt\hbox{$\scriptscriptstyle\smile$}}
\over H} }}}_{\rm{p}}})}} \le \dfrac{{c {\lambda _{\max }}({{\bf{R}}_{\rm{c}}} - {\varepsilon _{\rm{p}}}{{\bf{R}}_{\rm{s}}})}}{{{\varepsilon _{\rm{p}}}{\sigma ^2_{\rm{p}}}}}} \right\} \le 1 - \tau.
\end{array}
\end{align}
By applying the lemma in \cite[Lemma 1]{chisquare}, it follows
\begin{align}
\label{50011}
\begin{array}{l}
 {\lambda _{\max }}({{\bf{R}}_{\rm{c}}} - {\varepsilon _{\rm{p}}}{{\bf{R}}_{\rm{s}}}) \le \Phi _M^{ - 1}(1 - \tau )\dfrac{{{\varepsilon _{\rm{p}}}{\sigma ^2_{\rm{p}}}}}{c}.
\end{array}
\end{align}
This thus completes the proof.
\section*{Appendix B}                            
We now prove that $\tilde \Xi $ is the globally optimal solution to the optimization problem P2. Because the optimization objective $L({\bf{R}},{\delta _1})$ in (\ref{23}) is determined by the covariance matrix ${\bf{R}}$ of the transmitted signal, we only need to verify that $\tilde \Xi $ is the viable solution to P3, i.e. $\tilde \Xi $ satisfies all the constraints in P3. Following \cite{liuxiang}, we know the radar waveform covariance matrix ${\bf{\tilde R}} - \sum\nolimits_{k = 1}^K {{{{\bf{\tilde R}}}_k}} $ is semi-definite Hermitian matrix, where ${\bf{\tilde R}}={\bf{\hat R}}$. Accordingly, we just need to verify the other three constraints ((\ref{20})--(\ref{22})).

First, we can derive that
\begin{align}
\label{51}
\begin{array}{l}
{\bf{h}}_{k}^{\rm{H}}{{{\bf{\tilde R}}}_{k}}{{\bf{h}}_{k}}{\rm{ = }}{\bf{h}}_{k}^{\rm{H}}{{{\bf{\tilde w}}}_{k}}{\bf{\tilde w}}_{k}^{\rm{H}}{{\bf{h}}_{k}}\\
{\rm{ = }}{\bf{h}}_{k}^{\rm{H}}{\left( {{\bf{h}}_{k}^{\rm{H}}{{{\bf{\hat R}}}_{k}}{{\bf{h}}_{k}}} \right)^{ - 1/2}}{{{\bf{\hat R}}}_{k}}{{\bf{h}}_{k}} {\left( {{{\left( {{\bf{h}}_{k}^{\rm{H}}{{{\bf{\hat R}}}_{k}}{{\bf{h}}_{k}}} \right)}^{ - 1/2}}{{{\bf{\hat R}}}_{k}}{{\bf{h}}_{k}}} \right)^{\rm{H}}}{{\bf{h}}_{k}}\\
{\rm{ = }}{\left( {{\bf{h}}_{k}^{\rm{H}}{{{\bf{\hat R}}}_{k}}{{\bf{h}}_{k}}} \right)^{ - 1}}{\bf{h}}_{k}^{\rm{H}}{{{\bf{\hat R}}}_{k}}{{\bf{h}}_{k}}{\bf{h}}_{k}^{\rm{H}}{{{\bf{\hat R}}}_{k}}{{\bf{h}}_{k}}\\
{\rm{ = }}{\bf{h}}_{k}^{\rm{H}}{{{\bf{\hat R}}}_{k}}{{\bf{h}}_{k}}.
\end{array}
\end{align}
Utilizing (\ref{51}) and (\ref{20}), we obtain the inequality as
\begin{align}
\label{52}
\begin{array}{*{20}{l}}
	{\left( {1{\rm{ + }}\dfrac{1}{{\varepsilon _{{\rm{u}},k}^{}}}} \right){\bf{h}}_k^{\rm{H}}{{{\bf{\tilde R}}}_k}{{\bf{h}}_k} = \left( {1{\rm{ + }}\dfrac{1}{{\varepsilon _{{\rm{u}},k}^{}}}} \right){\bf{h}}_k^{\rm{H}}{{{\bf{\hat R}}}_k}{{\bf{h}}_k}}\\
	{ \ge {\bf{h}}_k^{\rm{H}}{\bf{\hat R}}{{\bf{h}}_k} + {P_{\rm{a}}}|{h_{{\rm{a}},k}}{|^2} + \sigma _{\rm{c}}^2}\\
	{ = {\bf{h}}_k^{\rm{H}}{\bf{\tilde R}}{{\bf{h}}_k} + {P_{\rm{a}}}|{h_{{\rm{a}},k}}{|^2} + \sigma _{\rm{c}}^2.}
\end{array}
\end{align}
For the SINR constraint of AE, i.e. (\ref{21}), it follows
\begin{align}
\label{53}
\begin{array}{l}
{\bf{a}}{^{\rm{H}}(\theta_Q )}{{\bf{\tilde R}}_k}{\bf{a}}(\theta_Q ) = {\left( {{\bf{h}}_{k}^{\rm{H}}{{{\bf{\hat R}}}_k}{{\bf{h}}_{k}}} \right)^{ - 1}}{\left| {{\bf{a}}{{(\theta_Q )}^{\rm{H}}}{{{\bf{\hat R}}}_k}{{\bf{h}}_{k}}} \right|^2}\\
\mathop  \le \limits^{\rm{(a)}} {\left( {{\bf{h}}_{k}^{\rm{H}}{{{\bf{\hat R}}}_k}{{\bf{h}}_{k}}} \right)^{ - 1}} \left( {{\bf{h}}_{k}^{\rm{H}}{{{\bf{\hat R}}}_k}{{\bf{h}}_{k}}} \right)\left( {{\bf{a}}{{(\theta_Q )}^{\rm{H}}}{{{\bf{\hat R}}}_k}{\bf{a}}(\theta_Q )} \right)\\
= {\bf{a}}{^{\rm{H}}(\theta_Q )}{\bf{\hat R_{{\it{k}}}a}}(\theta_Q ),
\end{array}
\end{align}
where (a) is due to Cauchy-Schwartz inequality \cite{19}. Then, substituting (\ref{53}) into (\ref{21}), we calculate
\begin{align}
\label{55}
\begin{array}{l}
{\bf{a}}{^{\rm{H}}(\theta_Q )}{\bf{\tilde Ra}}(\theta_Q ) + \dfrac{{{\sigma ^2_{\rm{s}}}}}{{|\beta_Q {|^2}}} = {\bf{a}}{^{\rm{H}}(\theta_Q )}{\bf{\hat Ra}}(\theta_Q ) + \dfrac{{{\sigma ^2_{\rm{s}}}}}{{|\beta_Q {|^2}}}\\
 \ge \left( {1{\rm{ + }}{\dfrac{1}{{\varepsilon _{{\rm{a}}}^{}}}}} \right){\bf{a}}{^{\rm{H}}(\theta_Q )}\sum\nolimits_{k = 1}^K {{{{\bf{\hat R}}}_k}} {\bf{a}}(\theta_Q )\\
 \ge \left( {1{\rm{ + }}{\dfrac{1}{{\varepsilon _{{\rm{a}}}^{}}}}} \right){\bf{a}}{^{\rm{H}}(\theta_Q )}\sum\nolimits_{k = 1}^K {{{{\bf{\tilde R}}}_k}} {\bf{a}}(\theta_Q ).
\end{array}
\end{align}
Similarly, we need to verify that $\tilde \Xi $ satisfies the SINR constraint of PE, i.e. (\ref{19}), and can show
\begin{align}
\label{56}
\begin{array}{l}
{\bf{h}}_{\rm{p}}^{\rm{H}}{{{\bf{\tilde R}}}_k}{{\bf{h}}_{\rm{p}}} = {\bf{h}}_{\rm{p}}^{\rm{H}}{{{\bf{\tilde w}}}_k}{\bf{\tilde w}}_k^{\rm{H}}{{\bf{h}}_{\rm{p}}}\\
 = {\left( {{\bf{h}}_{k}^{\rm{H}}{{{\bf{\hat R}}}_k}{{\bf{h}}_{k}}} \right)^{ - 1}}{\bf{h}}_{\rm{p}}^{\rm{H}}{{{\bf{\hat R}}}_k}{{\bf{h}}_{k}}{\bf{h}}_{k}^{\rm{H}}{{{\bf{\hat R}}}_k}{{\bf{h}}_{\rm{p}}}\\
 = {\left( {{\bf{h}}_{k}^{\rm{H}}{{{\bf{\hat R}}}_k}{{\bf{h}}_{k}}} \right)^{ - 1}}{\left| {{\bf{h}}_{\rm{p}}^{\rm{H}}{{{\bf{\hat R}}}_k}{{\bf{h}}_{k}}} \right|^2}.
\end{array}
\end{align}
By utilizing Cauchy-Schwartz inequality, we can compute
\begin{align}
\label{57}
{\left| {{\bf{h}}_{\rm{p}}^{\rm{H}}{{{\bf{\hat R}}}_k}{{\bf{h}}_{k}}} \right|^2} \le  {{\bf{h}}_{k}^{\rm{H}}{{{\bf{\hat R}}}_k}{{\bf{h}}_{k}}}  {{\bf{h}}_{\rm{p}}^{\rm{H}}{{{\bf{\hat R}}}_k}{{\bf{h}}_{\rm{p}}}} .
\end{align}
Then, substituting (\ref{56}) into (\ref{19}), we have
\begin{align}
\label{58}
\begin{array}{*{20}{l}}
{\lambda _{\max }}\left( {\left( {1 + {\varepsilon _{\rm{p}}}} \right)\sum\nolimits_{k = 1}^K {{{\bf{\tilde R}}_k}}  - {\varepsilon _{\rm{p}}}{\bf{\tilde R}}} \right) \le \Phi _M^{ - 1}\left( {1 - \tau } \right){\varepsilon _{\rm{p}}}{\sigma ^2_{\rm{p}}.}
\end{array}
\end{align}
This thus completes the proof.
\section*{Appendix C}                            
The proof process is divided into three parts. In the first part, we will prove that the radar covariance matrix ${{\bf{\tilde R}} - {{\bf{\tilde W}}_{\rm{c}}}{\bf{\tilde W}}_{\rm{c}}^{\rm{H}}}$ is a semi-definite Hermitian matrix, which can be expanded as
    \begin{align}
\label{59}
\begin{array}{*{20}{l}}
{{\bf{\tilde R}} - {{{\bf{\tilde W}}}_{\rm{c}}}{\bf{\tilde W}}_{\rm{c}}^{\rm{H}}}\\
{ = {\bf{\tilde R}} - {{{\bf{\tilde R}}}_{\rm{c}}} + {{{\bf{\tilde R}}}_{\rm{c}}} - {{{\bf{\tilde W}}}_{\rm{c}}}{\bf{\tilde W}}_{\rm{c}}^{\rm{H}}}\\
{ = {\bf{\tilde R}} - {{{\bf{\tilde R}}}_{\rm{c}}} + {\bf{D}}\left( {{\bf{I}} - {\bf{\tilde U}}{{{\bf{\tilde U}}}^{\rm{H}}}} \right){{\bf{D}}^{\rm{H}}}.}
\end{array}
\end{align}
Here, ${\bf{\tilde R}} - {{\bf{\tilde R}}_{\rm{c}}}$ is semi-definite and it can be decomposed by the Cholesky decomposition. Since ${{\bf{\tilde U}}}$ is the sub-matrix containing the first {\it{K}} columns of unitary matrix, ${{\bf{I}} - {{\bf{\tilde U}}{{{\bf{\tilde U}}}^{\rm{H}}}}}$ is a positive semi-definite matrix. Thereby the last term of (\ref{59}) is also positive semi-definite.

In the second part, we aim to prove that ${\bf{\tilde W}}_{\rm{c}}$ and ${\bf{\tilde W}}_{\rm{s}}$ satisfy the ZF constraints. According to ((28), (29)), we compute
\begin{align}
\label{60}
{{\bf{H}}_{\rm{u}}}{{\bf{\tilde R}}_{\rm{c}}}{\bf{H}}_{\rm{u}}^{\rm{H}} = {{\bf{H}}_{\rm{u}}}{\bf{D}}{{\bf{D}}^{\rm{H}}}{\bf{H}}_{\rm{u}}^{\rm{H}} = {{\bf{B}}_{\rm{L}}}{\bf{B}}_{\rm{L}}^{\rm{H}} = {\rm{diag}}\left( {{\rho _1},...,{\rho _K}}  \right).
\end{align}
Noting that ${{\bf{B}}_{\rm{L}}}{\bf{B}}_{\rm{L}}^{\rm{H}}$ is the Cholesky decompositions of the matrix ${\rm{diag}}\left( {{\rho _1},...,{\rho _K}}  \right)$, it follows
\begin{align}
\label{61}
\begin{array}{*{20}{l}}
{{{\bf{H}}_{\rm{u}}}{{{\bf{\tilde W}}}_{\rm{c}}} = {{\bf{H}}_{\rm{u}}}{\bf{D\tilde U}}}
{ = \left[ {{{\bf{B}}_{\rm{L}}},{{\bf{0}}_{K \times \left( {M - K} \right)}}} \right]{{\bf{U}}_2}{\bf{\tilde U}}}
{ = {{\bf{B}}_{\rm{L}}}.}
\end{array}
\end{align}
Thus, ${{\bf{H}}_{\rm{u}}}{{\bf{\tilde W}}_{\rm{c}}}{\bf{\tilde W}}_{\rm{c}}^{\rm{H}}{\bf{H}}_{\rm{u}}^{\rm{H}}={\rm{diag}}\left( {{\rho _1},...,{\rho _K}}  \right)$ is satisfied. Moreover, for the sensing beamforming matrix, we arrive at
\begin{align}
\label{62}
\begin{array}{*{20}{l}}
{{{\bf{H}}_{\rm{u}}}{{\bf{\tilde W}}_{\rm{s}}}{\bf{\tilde W}}_{\rm{s}}^{\rm{H}}{\bf{H}}_{\rm{u}}^{\rm{H}} = {{\bf{H}}_{\rm{u}}}\left( {{\bf{\tilde R}} - {{\bf{\tilde W}}_{\rm{c}}}{\bf{\tilde W}}_{\rm{c}}^{\rm{H}}} \right){\bf{H}}_{\rm{u}}^{\rm{H}}}  = {\bf{0}}.
\end{array}
\end{align}
From this we can readily obtain ${\bf{H}}{{\bf{\tilde W}}_{\rm{s}}} = {\bf{0}}$.

In the third part, we will prove that ${\bf{\tilde W}}_{\rm{c}}$ and ${\bf{\tilde W}}_{\rm{s}}$ satisfy the two eavesdropping constraints((\ref{21}), (\ref{22})). According to the properties of the semi-definite matrix, for any nonzero vector {\bf{y}}, we can show
\begin{align}
\label{63}
{{\bf{y}}^{\rm{H}}}\left( {{\bf{I}} - {\bf{\tilde U}}{{{\bf{\tilde U}}}^{\rm{H}}}} \right){\bf{y}} \ge 0.
\end{align}
For the AE, let ${\bf{y}} = {\bf{D}}^{\rm{H}}{\bf{a}}(\theta_Q )$, we can calculate
\begin{align}
\label{64}
\begin{array}{*{20}{l}}
{{\bf{a}}{^{\rm{H}}(\theta_Q )}{\bf{D}}\left( {{\bf{I}} - {\bf{\tilde U}}{{{\bf{\tilde U}}}^{\rm{H}}}} \right){{\bf{D}}^{\rm{H}}}{\bf{a}}(\theta_Q )}\\
{ = {\bf{a}}{^{\rm{H}}(\theta_Q )}{{{\bf{\tilde R}}}_{\rm{c}}}{\bf{a}}(\theta_Q ) - {\bf{a}}{^{\rm{H}}(\theta_Q )}{{\bf{\tilde W}}_{\rm{c}}}{\bf{\tilde W}}_{\rm{c}}^{\rm{H}}{\bf{a}}(\theta_Q )}{ \ge 0}.
\end{array}
\end{align}
Then, substituting (\ref{64}) into (\ref{21}), we can obtain
\begin{align}
\label{65}
\begin{array}{l}
{\bf{a}}^{\rm{H}}{(\theta_Q )}{\bf{\tilde Ra}}(\theta_Q ) + \dfrac{{{\sigma _{\rm{s}}^2}}}{{|\beta_Q {|^2}}} \ge \left( {1{\rm{ + }}{\dfrac{1}{{\varepsilon _{{\rm{a}}}^{}}}}} \right){\bf{a}}^{\rm{H}}{(\theta_Q )}{{{\bf{\tilde R}}}_{\rm{c}}}{\bf{a}}(\theta_Q )\\
{\rm{                               }} \ge \left( {1{\rm{ + }}{\dfrac{1}{{\varepsilon _{{\rm{a}}}^{}}}}} \right){\bf{a}}^{\rm{H}}{(\theta_Q )}{{\bf{\tilde W}}_{\rm{c}}}{\bf{\tilde W}}_{\rm{c}}^{\rm{H}}{\bf{a}}(\theta_Q ).
\end{array}
\end{align}
The same goes for constraints on PE, letting ${\bf{y}} = {\bf{D}}^{\rm{H}}{{\bf{h}}_{\rm{p}}}$, and this yields
\begin{align}
\label{66}
\begin{array}{*{20}{l}}
{{\bf{h}}_{\rm{p}}^{\rm{H}}{\bf{D}}\left( {{\bf{I}} - {\bf{\tilde U}}{{{\bf{\tilde U}}}^{\rm{H}}}} \right){{\bf{D}}^{\rm{H}}}{{\bf{h}}_{\rm{p}}}}\\
{ = {\bf{h}}_{\rm{p}}^{\rm{H}}{{{\bf{\tilde R}}}_{\rm{c}}}{{\bf{h}}_{\rm{p}}} - {\bf{h}}_{\rm{p}}^{\rm{H}}{{\bf{\tilde W}}_{\rm{c}}}{\bf{\tilde W}}_{\rm{c}}^{\rm{H}}{{\bf{h}}_{\rm{p}}}}{ \ge 0}.
\end{array}
\end{align}
Substituting (\ref{66}) into PE constraint, we can obtain
\begin{align}
\label{67}
\begin{array}{l}
{\rm{Pr}}\left( {{{\bf{h}}_{\rm{p}}^{\rm{H}}}\left( {\left( {1 + {\varepsilon _{\rm{p}}}} \right){{\bf{\tilde W}}_{\rm{c}}}{\bf{\tilde W}}_{\rm{c}}^{\rm{H}} - {\varepsilon _{\rm{p}}}{\bf{\hat R}}} \right){{\bf{h}}_{\rm{p}}} \le {\varepsilon _{\rm{p}}}{\sigma ^2_{\rm{p}}}} \right)\\
 \ge {\rm{Pr}}\left( {{{\bf{h}}_{\rm{p}}^{\rm{H}}}\left( {\left( {1 + {\varepsilon _{\rm{p}}}} \right){{\bf{\tilde R}}}_{\rm{c}} - {\varepsilon _{\rm{p}}}{\bf{\tilde R}}} \right){{\bf{h}}_{\rm{p}}} \le {\varepsilon _{\rm{p}}}{\sigma ^2_{\rm{p}}}} \right)\ge \tau.
\end{array}
\end{align}
It can be observed that the constructed beamforming matrices ${\bf{\tilde W}}_{\rm{c}}$ and ${\bf{\tilde W}}_{\rm{s}}$ satisfy all constraints in (P4) and they are the optimal beamforming matrices of (P4).
\section*{Acknowledgments}
The authors are very grateful to the reviewers and the Editor, Dr. Musa Furkan Keskin, for their comments that improved the presentation of this paper. They also warmly thank Prof. Bin Dai for his interesting discussion and several exciting suggestions along this work.

\bibliographystyle{IEEEtran}
\bibliography{paper} 



\vfill

\par\noindent
\parbox[t]{\linewidth}{
\noindent\parpic{\includegraphics[height=1.2in,width=1in,clip,keepaspectratio]{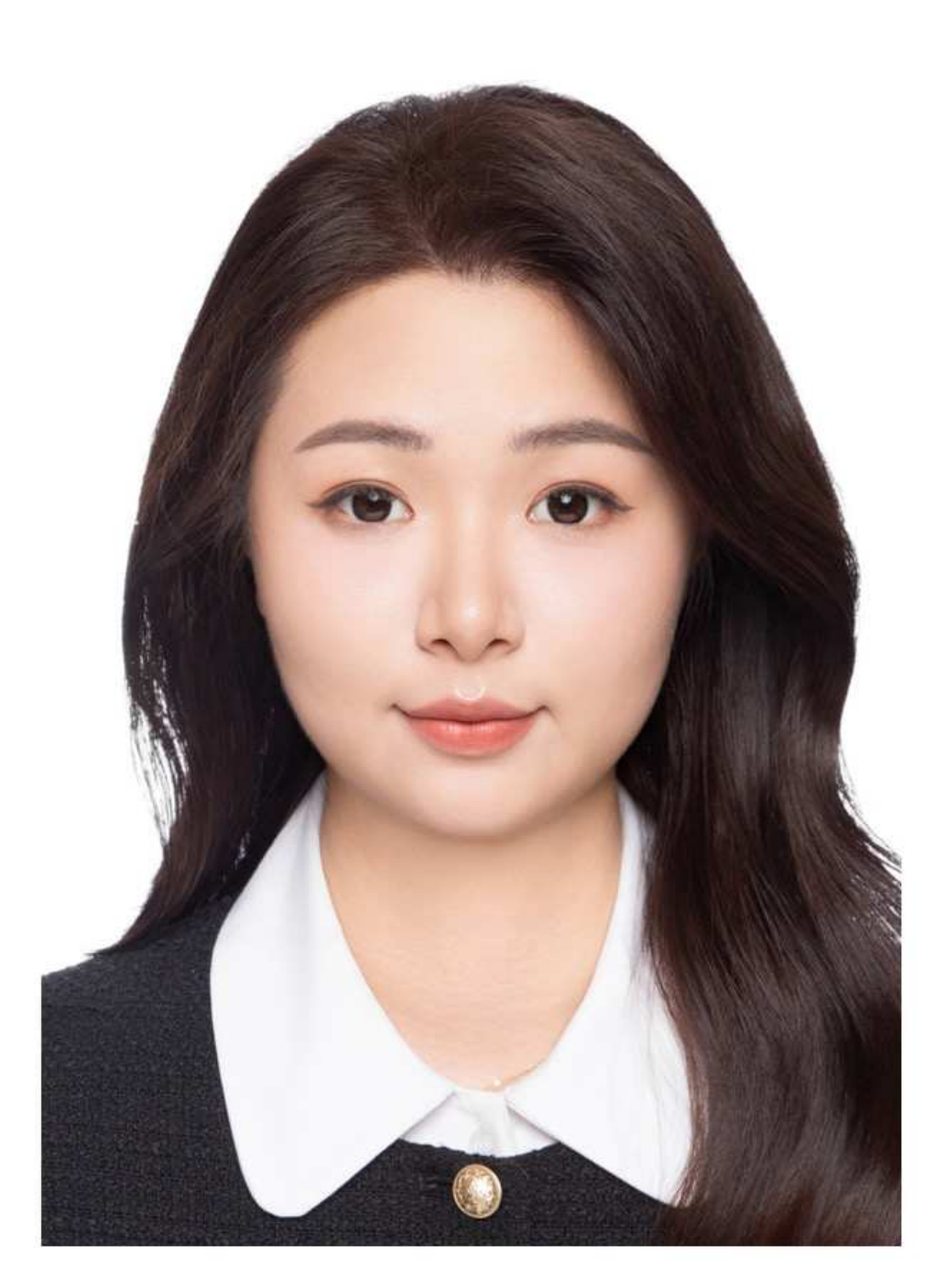}}
\noindent {\bf Meiding Liu}\
received the B.S. and M.S. degrees from Chongqing University of Posts and Telecommunications (CQUPT), Chongqing, China, in 2020 and 2023, respectively. She is currently pursuing the Ph.D. degree with the School of Information Science and Technology, Southwest Jiaotong University (SWJTU). Her research interests include integrated sensing and communication systems (ISAC), IRS-assisted communication, physical-layer security, radar signal processing, convex optimization.}
\vspace{1\baselineskip}

\par\noindent
\parbox[t]{\linewidth}{
\noindent\parpic{\includegraphics[height=1.2in,width=1in,clip,keepaspectratio]{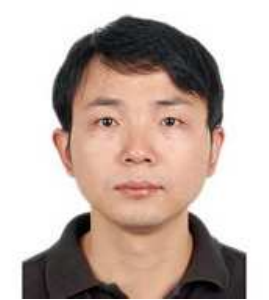}}
\noindent {\bf Zhengchun Zhou (Member, IEEE)}\
received the B.S. and M.S. degrees in mathematics and the Ph.D. degree in information security from Southwest Jiaotong University, Chengdu, China, in 2001, 2004, and 2010, respectively. From 2012 to 2013, he was a postdoctoral member in the Department of Computer Science and Engineering, the Hong Kong University of Science and Technology. From 2013 to 2014, he was a research associate in the Department of Computer Science and Engineering, the Hong Kong University of Science and Technology. Now he is a professor in the School of Information Science and Technology (and also a professor in the School of Mathematics), Southwest Jiaotong University. His research interests include coding theory, cryptography, intelligent information processing. He is an associated editor of several journals including IEEE Transactions on Cognitive Communications and Networking, Cryptography and Communications and Advances in Mathematics of Communications. Dr. Zhou was the recipient of the National excellent Doctoral Dissertation award in 2013 (China).}
\vspace{1\baselineskip}

\par\noindent
\parbox[t]{\linewidth}{
\noindent\parpic{\includegraphics[height=1.2in,width=1in,clip,keepaspectratio]{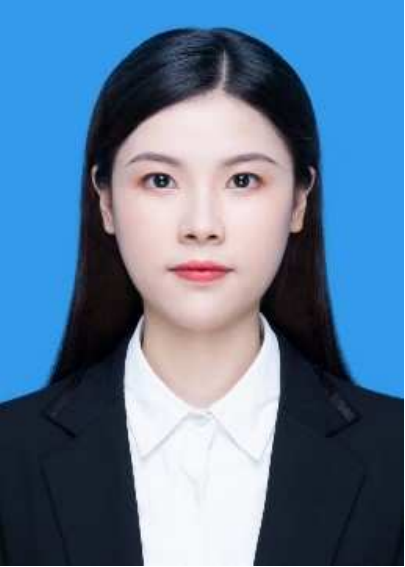}}
\noindent {\bf Qiao Shi (Member, IEEE)}\
received the B.S. degree in electronic and information engineering from the Southwest University of Science and Technology, Mianyang, China, in 2016, and the Ph.D. degree in signal and information processing from the University of Electronic Science and Technology of China (UESTC), Chengdu, China, in 2022. From 2021 to 2022, she was a Visiting Student with Communication and Intelligent System Laboratory, Korea University, Seoul, South Korea. She is currently a lecturer in the School of Information Science and Technology, Southwest Jiaotong University. Her research interests include radar signal processing and waveform design for integrated sensing and communication systems.}
\vspace{1\baselineskip}

\par\noindent
\parbox[t]{\linewidth}{
	\noindent\parpic{\includegraphics[height=1.2in,width=1in,clip,keepaspectratio]{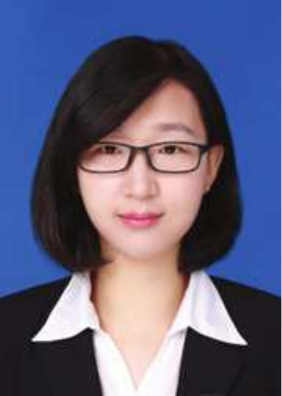}}
	\noindent {\bf Guyue Li (Member, IEEE)}\
	received the B.S. degree in information science and technology and the Ph.D. degree in information security from Southeast University, Nanjing, China, in 2011 and 2017, respectively. From June 2014 to August 2014, she was a Visiting Student with the Department of Electrical Engineering, Tampere University of Technology, Tampere, Finland. She is currently an Associate Professor with the School of Cyber Science and Engineering, Southeast University, and a Visiting Scholar with the Tampere University of Technology, and Université Gustave Eiffel (ESIEE PARIS), Noisy-le-Grand, France. Her current research interests include wireless network attacks, physical-layer security solutions for 5G and 6G, secret key generation, radio frequency fingerprints, and reconfigurable intelligent surfaces. She is currently serving as an Editor for IEEE Communication Letters and an Associate Editor for EURASIP Journal on Wireless Communications and Networking.}
\vspace{1\baselineskip}

\par\noindent
\parbox[t]{\linewidth}{
	\noindent\parpic{\includegraphics[height=1.2in,width=1in,clip,keepaspectratio]{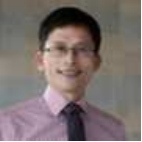}}
	\noindent {\bf Zilong Liu (Senior Member, IEEE)}\
	received the bachelor's degree in electronics and information engineering from Huazhong University of Science and Technology (HUST), Wuhan, China, the master's degree in electronic engineering from Tsinghua University, Beijing, China, and the Ph.D. degree in electrical and electronic engineering, Nanyang Technological University (NTU), Singapore, in 2004, 2007 and 2014, respectively. Since July 2008, he has been with the School of Electrical and Electronic Engineering, NTU, first as a Research Associate and since November 2014, as a Research Fellow. He was a Visitor at the University of Melbourne (UoM), Parkville, Vic., Australia, from May 2012 to February 2013 (hosted by prof. Udaya Parampalli), and a Visiting Ph.D. Student at the Hong Kong University of Science and Technology (HKUST) from June 2013 to July 2013 (hosted by prof. Wai Ho Mow). He is currently an Associate Professor with the School of Computer Science and Electronic Engineering, University of Essex. His research lies in the interplay of coding, signal processing, and communications, with a major objective of bridging theory and practice. Recently, he has developed an interest in advanced 6G V2X communication, sensing, and localization technologies for future connected autonomous vehicles as well as machine learning for enhanced communications and networking.}
\vspace{1\baselineskip}

\par\noindent
\parbox[t]{\linewidth}{
	\noindent\parpic{\includegraphics[height=1.2in,width=1in,clip,keepaspectratio]{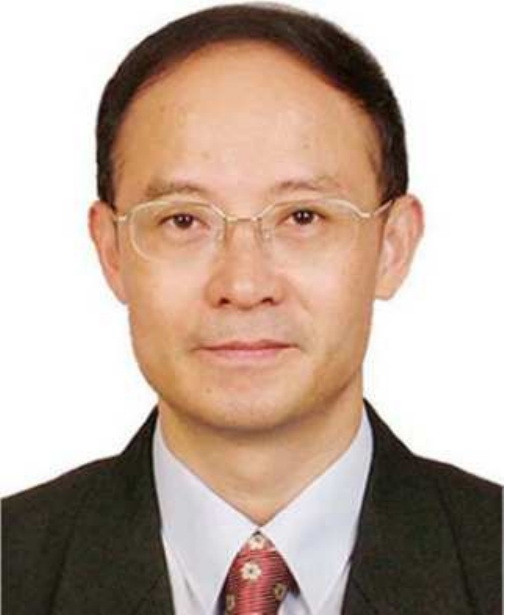}}
	\noindent {\bf Pingzhi Fan (Fellow, IEEE)}\
	received the M.Sc. degree in computer science from the Southwest Jiaotong University, China, in 1987, and the Ph.D. degree in electronic engineering from Hull University, U.K., in 1994. He is currently a Distinguished Professor and Director of the Institute of Mobile Communications, Southwest Jiaotong University, China, and a Visiting Professor of Leeds University, UK (1997-), a Guest Professor of Shanghai Jiaotong University (1999-). He is a recipient of the UK ORS Award (1992), the NSFC Outstanding Young Scientist Award (1998), IEEE VTS Jack Neubauer Memorial Award (2018), and 2018 IEEE SPS Signal Processing Letters Best Paper Award. His current research interests include vehicular communications, massive multiple access and coding techniques, etc. He served as General Chair or TPC Chair of a number of international conferences including VTC 2016 Spring, IWSDA 2019, ITW 2018 etc. He is the Founding Chair of IEEE Chengdu (CD) Section, IEEE VTS BJ Chapter and IEEE ComSoc CD Chapter. He also served as an EXCOM member of IEEE Region 10, IET(IEE) Council and IET Asia Pacific Region. He has published over 300 international journal papers and eight books (including edited), and is the Inventor of 25 granted patents. He is an IEEE VTS Distinguished Lecturer (2015-2019), and a fellow of IET, CIE, and CIC.}
\vspace{1\baselineskip}

\par\noindent
\parbox[t]{\linewidth}{
	\noindent\parpic{\includegraphics[height=1.2in,width=1in,clip,keepaspectratio]{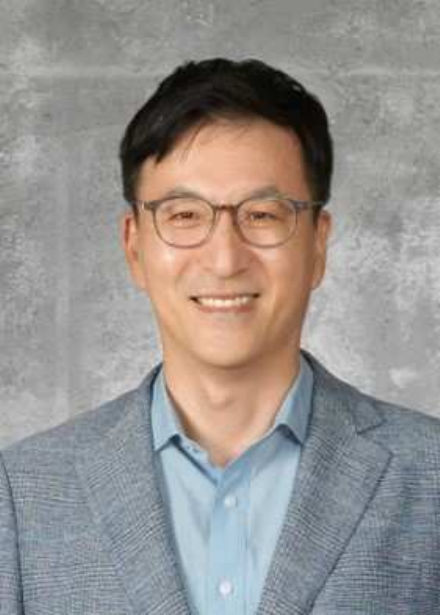}}
	\noindent {\bf Inkyu Lee (Fellow, IEEE)}\
	received the B.S. degree (Hons.) in control and instrumentation engineering from Seoul National University, Seoul, South Korea, in 1990, and the M.S. and Ph.D. degrees in electrical engineering from Stanford University, Stanford, CA, USA, in 1992 and 1995, respectively. From 1995 to 2002, he was a member of the Technical Staff with Bell Laboratories, Lucent Technologies, Murray Hill, NJ, USA, where he studied high-speed wireless system designs. Since 2002, he has been with Korea University, Seoul, where he is currently a Professor with the School of Electrical Engineering. He has also served as the Department Head of the School of Electrical Engineering, Korea University, from 2019 to 2021. In 2009, he was a Visiting Professor with the University of Southern California, Los Angeles, CA, USA. He has authored or co-authored more than 230 journal articles in IEEE publications and holds 30 U.S. patents granted or pending. His research interests include digital communications and signal processing techniques applied for next-generation wireless systems. Dr. Lee was a recipient of the IT Young Engineer Award from the IEEE/IEEK Joint Award in 2006, the Best Paper Award from the IEEE Vehicular Technology Conference in 2009, the Best Young Engineer Award from the National Academy of Engineering of Korea in 2013, and the Korea Engineering Award from the National Research Foundation of Korea in 2017. He served as an Associate Editor for the IEEE TRANSACTIONS ON COMMUNICATIONS from 2001 to 2011 and the IEEE TRANSACTIONS ON WIRELESS COMMUNICATIONS from 2007 to 2011. He was a Chief Guest Editor of the IEEE JOURNAL ON SELECTED AREAS IN COMMUNICATIONS Special Issue on \enquote{4G wireless systems} in 2006. He was a TPC Co-Chair for IEEE International Conference on Communications in 2022. He also served as the Co-Editor-in-Chief for the Journal of Communications and Networks from 2019 to 2024. He was elected as a member of the National Academy of Engineering of Korea in 2015. He is the Director of \enquote{Augmented Cognition Meta-Communication} ERC Research Center awarded from the National Research Foundation of Korea. He serves as a president-elect for the Korean Institute of Communications and Information Sciences (KICS).  He is a Distinguished Lecturer of IEEE.}
\vspace{1\baselineskip}

\end{document}